\documentclass[prb,aps,twocolumn,superscriptaddress,showpacs,footinbib]{revtex4-1}
\usepackage[colorlinks=true,linkcolor=black, citecolor=blue, urlcolor=blue, 
    unicode=true]{hyperref}

\usepackage{graphicx,amsmath,braket,tikz,amssymb,footmisc,lipsum,array,tabularx,float,comment}

\usepackage{cleveref}

\crefname{equation}{}{}
\Crefname{equation}{}{}
\crefrangelabelformat{equation}{#3#1#4--#5#2#6}

\crefmultiformat{equation}{#2#1#3}{, #2#1#3}{#2#1#3}{#2#1#3}
\Crefmultiformat{equation}{#2#1#3}{, #2#1#3}{#2#1#3}{#2#1#3}

\allowdisplaybreaks
\begin{document}

\title{\texorpdfstring{Excitonic transverse and amplitude fluctuations in the noncollinear and charge-ordered RbFe$^{2+}$Fe$^{3+}$F$_{6}$}{}}

\author{H. Lane}
\affiliation{School of Physics and Astronomy, University of Edinburgh, Edinburgh EH9 3JZ, UK}
\affiliation{School of Chemistry and Centre for Science at Extreme Conditions, University of Edinburgh, Edinburgh EH9 3FJ, UK}
\affiliation{ISIS Pulsed Neutron and Muon Source, STFC Rutherford Appleton Laboratory, Harwell Campus, Didcot, Oxon, OX11 0QX, United Kingdom}
\author{M. Songvilay}
\affiliation{Institut N{\'e}el, CNRS and Universit{\'e} Grenoble Alpes, 38000 Grenoble, France}
\author{R. A. Ewings}
\affiliation{ISIS Pulsed Neutron and Muon Source, STFC Rutherford Appleton Laboratory, Harwell Campus, Didcot, Oxon, OX11 0QX, United Kingdom}
\author{C. Stock}
\affiliation{School of Physics and Astronomy, University of Edinburgh, Edinburgh EH9 3JZ, UK}

\date{\today}

\begin{abstract}

RbFe$^{2+}$Fe$^{3+}$F$_{6}$ is an example of an antiferromagnet with charge ordering of the octahedrally coordinated Fe$^{2+}$ and Fe$^{3+}$ ions.  As well as different spin values, Fe$^{2+}$ ($S=2$) and Fe$^{3+}$ ($S={5\over2}$) possess differing orbital ground states with Fe$^{2+}$ having an orbital degeneracy with an effective orbital angular momentum of $l=1$.  The resulting low temperature magnetic structure is non collinear with the spins aligned perpendicular to nearest neighbors (S. W. Kim \textit{et al.} Chem. Sci. {\bf{3}}, 741 (2012)).  The combination of an orbital degeneracy and non collinear spin arrangements introduces the possibility for unusual types of excitations such as amplitude modes of the order parameter.  In this paper we investigate this by applying a multi-level analysis to model neutron spectroscopy data (M. Songvilay \textit{et al.} Phys. Rev. Lett. {\bf{121}}, 087201 (2018)). In particular, we discuss the possible origins of the momentum and energy broadened continuum scattering observed in terms of amplitude fluctuations allowed through the presence of an orbital degree of freedom on the Fe$^{2+}$ site.  We extend previous spin-orbit exciton models based on a collinear spin structure to understand the measured low-energy excitations and also to predict and discuss possible amplitude mode scattering in RbFe$^{2+}$Fe$^{3+}$F$_{6}$.

\end{abstract}


\maketitle

\section{Introduction}

The concept of spin waves was first introduced by Bloch to describe the renormalization of the spontaneous magnetization of the simple ferromagnet \cite{Bloch30:61}. Since this initial work, subsequent contributions by Dyson~\cite{Dyson56:102}, and Holstein and Primakoff~\cite{Holstein40:58} have further expanded our understanding of the quasi-particle spectrum in magnetically ordered insulators. The importance of spin wave theory expanded significantly towards the latter part of the 20$^{\mathrm{th}}$ century, following the advent of neutron scattering techniques by Shull and Brockhouse which offered a way of directly probing the fundamental spin wave excitations of magnetic systems through the spin-spin correlation function. To this day, linear spin wave theory (LSWT) remains one of the primary means of investigating long range magnetically ordered phases of matter as a means of understanding the underlying interactions.  Its success in understanding spin interactions in insulators has resulted in several widely used computer routines for modelling neutron scattering data including SpinWave~\cite{Petit10:10,Poienar10:81}, SpinWaveGenie~\cite{Hahn19:Software}, and SpinW~\cite{Toth15:27}.  Such programmes have opened up neutron scattering to a broader user based and have contributed significantly to the success of new neutron instrumentation and the expansion of the user community.

LSWT is fundamentally a semiclassical technique and results from the expansion in $1/S$ about a classical ground state. Physically, it can be interpreted as describing transverse fluctuations of an ordered magnetic moment around a fixed direction.  It therefore enjoys greatest success in describing large-$S$ systems, where corrections to the leading order theory are small and the ground state is not dominated by quantum fluctuations~\cite{Songvilay126:21}. For small-$S$ systems there exist many fundamental excitations which are not well-described by LSWT such as spinons, breathers, and solitons \cite{Lake05:4,Mourigal13:9,Umegaki15:92}. Nonetheless, LWST has been surprisingly successful in describing physics away from the large-$S$, long-range ordered limit~\cite{Zhang19:122,Macdougal18:98,Pregelj18:98,Boldrin18:121}.

The typical LSWT treatment of coupled magnetic ions directly treats the spin degree of freedom based on a Hamiltonian with dominant Heisenberg terms. The effect of single-ion terms, such as spin-orbit coupling, in the magnetic Hamiltonian can be included perturbatively via Dzyaloshinskii-Moriya interactions and anisotropy terms.~\cite{Yosida:book} However, this treatment precludes the possibility of longitudinal amplitude fluctuations of the order parameter~\cite{Zhu19:1,Su20:102} which give rise to new types of excitations given that the observable $\hat{S}_{z}$ does not commute with the magnetic Hamiltonian.  Furthermore, the integrating out of the orbital degree of freedom can leave behind the incorrect single-ion ground state given the mixing of orbital and spin degrees of freedom.

Recently, effects of spin-orbit coupling on the magnetic excitations have been of intense interest in $4d$ or $5d$ transition metal ions~\cite{Rau16:7,HwanChun15:11,Banerjee17:356}.  However, given that the spin-orbit coupling scales as the atomic number squared ($\lambda \sim Z^{2}$)~\cite{Landau:book}, the energy scale for spin-orbit coupling is reduced for $3d$ transition metal ions, introducing the possibility of mixing of spin and orbital degrees of freedom on a energy scale measurable with neutron scattering~\cite{Sarte18:98_2}.  In such a situation, treatment of the magnetic excitations needs to incorporate the single-ion properties of the local crystalline electric field which define the eigenstates of the magnetic ions of interest.  

In this paper we revisit the spin excitations previously reported in RbFe$^{2+}$Fe$^{3+}$F$_{6}$~\cite{Kim11:3,Songvilay18:121}.  RbFe$^{2+}$Fe$^{3+}$F$_{6}$ has a structure related to the $\alpha$ pyrochlores $A_{2}B_{2}X_{6}X'$, but with a vacancy on one out of two $A$ cations and another on the $X'$ anion site that does not contribute to the $BX_{6}$ octahedra. Several compounds with similar structures have been reported in the literature.~\cite{Klepov21:143,Fennell19:15}  Charge order originates from the two different iron sites which have differing valences of Fe$^{2+}$ and Fe$^{3+}$.  While the magnetic ground state of Fe$^{3+}$ is $S={5\over 2}$ with each of the five $d$-orbitals half filled following Hund's rules and the Pauli principle, the situation for Fe$^{2+}$ is slightly more complicated with an extra electron occupying one of the $t_{2g}$ states with $S=2$ and an effective $l$=1 (which we discuss in more detail below).  As a direct result of this orbital degeneracy~\cite{Abragam:book,Gorev2016,Molokeev2013}, the Fe$^{2+}$F$_{6}$ octahedra are considerably more distorted than the Fe$^{3+}$F$_{6}$ octahedra.  

The goal of this paper is to investigate the spin fluctuations in RbFe$^{2+}$Fe$^{3+}$F$_{6}$, specifically the role of orbital contributions, which are coupled to the spin response via spin-orbit coupling, in the neutron cross section.  To understand the spin fluctuations and the role of the differing spin and orbital contributions from each of the iron sites, we present an extension to the Green's function formalism treating coupled multi-level sites to account for noncollinear magnetic order. We apply this formalism to the noncollinear charge-ordered antiferromagnet RbFe$^{2+}$Fe$^{3+}$F$_{6}$, calculating the excitation spectrum.  We discuss the low energy excitations and compare the results to previous neutron experiments and then investigate the amplitude fluctuations in the amplitude of the order parameter resulting from the nonconservation of $\hat{S}_{z}$.  

In this paper, we leverage the Green's function approach with the local symmetry to predict the existance of amplitude fluctuations of the ordered magnetic moment proportional to $\langle \hat{S}_{z} \rangle$.  The existence of this mode originates from the importance of spin-orbit coupling ($\propto \bf{L} \cdot \bf{S}$) in the magnetic Hamiltonian.  This additional terms means that the observable operator $\hat{S}_{z}$ no long commutes with the magnetic Hamiltonian and therefore fluctuations $\propto {{d \langle \hat{S}_{z} \rangle} \over {dt}}\neq 0$.  We show how the energy scale of the amplitude mode is controlled through single-ion terms in the Hamiltonian such as uniaxial anisotropy as well as spin-orbit coupling.  Owing to the single-ion nature of this excitation it is expected to be less dispersive than lower energy transverse excitations.

This manuscript is divided into five sections including this introduction.  In Section II we write out the definitions of the Green's function formalism applied here and extend it from previous works to a noncollinear magnet relevant here.  This section illustrates the role of single-ion physics in modelling neutron spectra and magnetic fluctuations.  In Section III we apply this to the situation in RbFe$^{2+}$Fe$^{3+}$F$_{6}$ and discuss the single-ion physics for Fe$^{2+}$ and Fe$^{3+}$ relevant in defining the ground state that is coupled via the Random Phase Approximation (RPA) in our Green's function approach.  In section IV we calculate the neutron response and and finish the paper with a discussion and concluding remarks in Section V.

\section{Green's function}\label{GreensFunction}

In this section we discuss the calculation of the Green's function in a noncollinear magnet and its relation to neutron spectroscopy. The Green's function formalism allows for the treatment of multi-level systems in a manner similar to SU(N) spin wave theory and the flavor wave expansion approach~\cite{Zhu19:1,Dong18:97,Hasegawa12:81,Elliot21:12,Bai21:17,Legros21:127}. By formulating the calculation in terms of response functions, a direct connection can be made to the neutron scattering intensity. 

\subsection{Relation to neutron spectroscopy}

The intensity measured with neutron scattering is directly proportional to the structure factor $S({\bf{q}}, \omega)$,

\begin{equation}
S({\bf{q}},\omega)=g_{L}^{2}f^2({\bf{q}})\sum_{\alpha \beta} (\delta_{\alpha \beta}-\hat{q}_{\alpha}\hat{q}_{\beta}) S^{\alpha \beta}({\bf{q}},\omega), 
\nonumber
\end{equation}

\noindent corresponding to a product of the Land\'{e} $g$-factor $g_{L}$, the magnetic form factor $f(\mathbf{q})$, a polarization factor providing sensitivity to the component perpendicular to the momentum transfer $\mathbf{q}$, and the dynamic spin structure factor $S^{\alpha \beta}({\bf{q}},\omega)$. This itself corresponds to the Fourier transform of the spin-spin correlations

\begin{equation}
S^{\alpha\beta}({\bf{q}},\omega)=\frac{1}{2\pi} \int dt e^{i\omega t} \langle \hat{S}^{\alpha} ({\bf{q}},t) \hat{S}^{\beta}(-{\bf{q}},0) \rangle,
\nonumber
\end{equation}

\noindent where $\alpha,\beta=x, y, z$. $S^{\alpha \beta}({\bf{q}},\omega)$ as written above considers only the spin contribution to the neutron scattering cross section.  The contributions from orbital fluctuations are ignored given that the expectation value of the orbital angular momentum $\langle {\bf{L}}\rangle \equiv$ 0 \emph{via} quenching for $d$-orbitals~\cite{Yosida:book}. The assumption to only consider the spin part of the neutron cross section depends on our experiment remaining in a single $|L, m_{L} \rangle$ multiplet and this is justified given the energy scales under consideration. Total moment sum rule analysis confirms that the spin contribution to the scattering cross section is dominant in RbFe$^{2+}$Fe$^{3+}$F$_{6}$~\cite{Songvilay:preprint}. As discussed in the following sections, orbital contributions to the structure enter via the spin-orbit ($\mathbf{L}\cdot\mathbf{S}$) coupling term in the magnetic Hamiltonian.

\indent The relation of the structure factor $S^{\alpha \beta}({\bf{q}},\omega)$ to the response function is given by the fluctuation-dissipation theorem 

\begin{equation}
S^{\alpha \beta}({\bf{q}},\omega)=-\frac{1}{\pi} \frac{1}{1-\exp(\omega/k{\rm{_{B}}}T)} \Im{G^{\alpha \beta} (\bf{q},\omega)}, \nonumber
\label{eq:2}
\end{equation}

\noindent and allows the magnetic neutron cross section to be defined in terms of a Green's response function $G^{\alpha \beta}(\bf{q},\omega)$~\cite{Zubarev60:3}. Recognizing that the neutron response function is proportional to the temperature dependent Bose factor multiplied by the Fourier transform of the retarded Green's function shows that calculating the Green's function response provides a means of modelling the neutron response.

\subsection{Laboratory frame}\label{collinear}

Building on previous work \cite{Holden74:9,Buyers75:11,Sarte19:100,Lane21:104,Lane21:104_2}, we now extend the Green's function formalism to treat noncollinear magnetic structures of arbitrary unit cell size. We begin by defining the Green's function equation of motion in the laboratory frame
\begin{equation}
G^{\alpha\beta}_{\tilde{\gamma}\tilde{\gamma}'}(i'j',t)=-i\Theta(t)\langle [\hat{S}^{\alpha}_{i'\tilde{\gamma}}(t),\hat{S}^{\beta}_{j'\tilde{\gamma}'}(0)]\rangle
\nonumber
\end{equation}
\noindent The indices $\alpha$ and $\beta$ label the spatial components in Cartesian coordinates, whilst $\tilde{\gamma}$ and $\tilde{\gamma}'$ label the atom site within the unit cell. The labeling convention for the indices used throughout this paper are summarized in Table \ref{Table:indices}.  
\begin{table}[h]
\caption{\label{Table:indices} Summary of labeling convention for indices.}
\begin{ruledtabular}
\begin{tabular}{cc}
Index & Description \\ 
\hline
$\gamma$, $\gamma'$ & sites within unit cell  \\ 
$i$, $j$ & unit cell  \\ 
$\alpha$, $\beta$, $\mu$, $\nu$ & Cartesian coordinates  \\ 
\end{tabular}
\end{ruledtabular}
\end{table}  

Causality is enforced by the Heaviside step function $\Theta(t)$, which precludes negative values of $t$. Taking the derivative of both sides with respect to time and multiplying by a factor of $i$, one finds that
\begin{equation}
\begin{split}
\label{eqofmot}
i\partial_{t}G^{\alpha\beta}_{\tilde{\gamma}\tilde{\gamma}'}(i'j',\omega)=&\delta(t)\langle [\hat{S}^{\alpha}_{i'\tilde{\gamma}}(t),\hat{S}^{\beta}_{j'\tilde{\gamma}'}]\rangle\\& \qquad -i\Theta(t)\langle [i\partial_{t}\hat{S}^{\alpha}_{i'\tilde{\gamma}}(t),\hat{S}^{\beta}_{j'\tilde{\gamma}'}]\rangle.
\end{split}
\end{equation}
\noindent Taking advantage of the Heisenberg equation of motion, $i\partial_{t}\hat{S}_{i\gamma}^{\alpha}(t)=[\hat{S}_{i\gamma}^{\alpha}(t),\mathcal{H}]$, and performing a Fourier transform in time, Eqn. \ref{eqofmot} can be written as
\begin{equation}
\begin{split}
\omega G^{\alpha\beta}_{\tilde{\gamma}\tilde{\gamma}'}(i'j',\omega)=&\langle [\hat{S}^{\alpha}_{i'\tilde{\gamma}},\hat{S}^{\beta}_{j'\tilde{\gamma}'}]\rangle+\\& \qquad G_{\tilde{\gamma}\tilde{\gamma}'}([\hat{S}^{\alpha}_{i'\tilde{\gamma}},\mathcal{H}],\hat{S}^{\beta}_{j'\tilde{\gamma}'},\omega).
\end{split}
\label{eqofmot2}
\end{equation}
For a system which consists of coupled multi-level sites, one can separate the Hamiltonian into single-ion and inter-ion terms 
\begin{equation}
\mathcal{H}=\sum_{i\gamma}\mathcal{H}'(i,\gamma)+\mathcal{H}_{int},
\nonumber
\end{equation}
\noindent where $\mathcal{H}'(i,\gamma)$ contains all of the manifestly single-ion terms such as spin-orbit coupling and the crystalline electric field whilst $\mathcal{H}_{int}$ describes the inter-ion terms such as the exchange interaction between sites and is therefore a sum over all bonds. In order that we expand about the correct single-ion ground state, we now perform a mean field decoupling $\mathbf{S}_{i\gamma}\to\langle\mathbf{S}_{i\gamma}\rangle +\delta\mathbf{S}_{i\gamma}$, discarding terms $\sim \mathcal{O}(\delta\mathbf{S}_{i\gamma})^{2}$. Following this decoupling, the single-ion Hamiltonian gains a molecular mean field Zeeman term which breaks spin-rotational symmetry. 

Assuming an interaction Hamiltonian of the form $\mathcal{H}_{int}=\frac{1}{2}\sum_{ij}^{\gamma\gamma'}\mathcal{J}_{ij}^{\gamma\gamma'}\mathbf{S}_{i\gamma}\cdot\mathbf{S}_{j\gamma'}$, where $\mathcal{J}_{ij}^{\gamma\gamma'}$ is a Heisenberg exchange parameter, the decoupled Hamiltonian becomes 
\begin{align*}
\mathcal{H}=&\mathcal{H}_{1}+\mathcal{H}_{2}\\
\mathcal{H}_{1}=&\sum_{i\gamma}\{\mathcal{H}'(i,\gamma)+ \sum_{j\gamma'}\mathcal{J}_{ij}^{\gamma\gamma'}[\mathbf{S}_{i\gamma}-\frac{1}{2}\langle \mathbf{S}_{i\gamma}\rangle]\langle \mathbf{S}_{j\gamma'}\rangle\}\\
\begin{split}
\mathcal{H}_{2}=&\frac{1}{2}\sum_{ij}^{\gamma\gamma'}\mathcal{J}_{ij}^{\gamma\gamma'}\mathbf{S}_{i\gamma}\cdot\mathbf{S}_{j\gamma'}\\&\qquad-\sum_{ij}^{\gamma\gamma'}\mathcal{J}_{ij}^{\gamma\gamma'}[\mathbf{S}_{i\gamma}-\frac{1}{2}\langle \mathbf{S}_{i\gamma}\rangle]\langle \mathbf{S}_{j\gamma'}\rangle
\end{split}.
\end{align*}
 
\noindent The projection of the spin operators onto the space spanned by the eigenvectors of the single-ion Hamiltonian, $\mathcal{H}_{1}$ can be written as
\begin{equation}
\label{spin}
\hat{S}^{\alpha}_{i\gamma}=\sum_{pq}S^{\gamma}_{\alpha pq}c^{\dagger}_{p}(i,\gamma)c_{q}(i,\gamma),
\nonumber
\end{equation}   
\noindent where the sum extends over all eigenstates, $\ket{p}$, of the Hamiltonian and $S^{\gamma}_{\alpha pq}=\bra{p}\hat{S}_{\gamma}^{\alpha}\ket{q}$. The operators $c_{q}^{\dagger}$ create the single-ion eigenstate $\ket{q}$. Now one must calculate the commutator in the right hand side of Eqn. \ref{eqofmot2} using the projected spin operator. The terms in the commutator are quartic in bosonic operators, however a random phase decoupling~\cite{Cooke73:7} can be performed, 

\begin{equation}
\begin{split}
&c^{\dagger}_{p}(i,\gamma)c_{q}(i,\gamma)c^{\dagger}_{m}(j,\gamma')c_{n}(j,\gamma')=\\&\qquad f_{p}(i,\gamma)\delta_{pq}c^{\dagger}_{m}(j,\gamma')c_{n}(j,\gamma')\\&\qquad +f_{m}(j,\gamma')\delta_{mn}c^{\dagger}_{p}(i,\gamma)c_{q}(i,\gamma),
\end{split}
\nonumber
\end{equation}

\noindent where $f_{p}(i,\gamma)$ is the Bose occupation factor of level $p$ on site $\gamma$ in unit cell $i$. In Cartesian coordinates, the commutator can be written as $[\hat{S}^{\alpha}_{i'\tilde{\gamma}},\mathcal{H}]=\sum_{s=1}^{4}\mathcal{C}_{s}$, with the individual terms given by,

\begin{subequations}
\begin{align}
\mathcal{C}_{1}=&\sum^{lkpq}_{j\gamma'}\phi_{qp}(i',\tilde{\gamma})c^{\dagger}_{k}(j,\gamma')c_{l}(j,\gamma')S_{\alpha qp}^{\tilde{\gamma}}S_{xpq}^{\tilde{\gamma}}S_{xkl}^{\gamma'}\mathcal{J}_{i'j}^{\tilde{\gamma}\gamma'}\label{commutatora}\\
\mathcal{C}_{2}=&\sum^{lkpq}_{j\gamma'}\phi_{qp}(i',\tilde{\gamma})c^{\dagger}_{k}(j,\gamma')c_{l}(j,\gamma')S_{\alpha qp}^{\tilde{\gamma}}S_{ypq}^{\tilde{\gamma}}S_{ykl}^{\gamma'}\mathcal{J}_{i'j}^{\tilde{\gamma}\gamma'}\label{commutatorb}\\
\mathcal{C}_{3}=&\sum^{lkpq}_{j\gamma'}\phi_{qp}(i',\tilde{\gamma})c^{\dagger}_{k}(j,\gamma')c_{l}(j,\gamma')S_{\alpha qp}^{\tilde{\gamma}}S_{zpq}^{\tilde{\gamma}}S_{zkl}^{\gamma'}\mathcal{J}_{i'j}^{\tilde{\gamma}\gamma'}\label{commutatorc}\\
\mathcal{C}_{4}=&\sum_{pq}\left(\omega_{p}-\omega_{q}\right)c^{\dagger}_{q}(i',\tilde{\gamma})c_{p}(i',\tilde{\gamma})S_{\alpha qp}^{\tilde{\gamma}},
\label{commutatord}
\end{align}
\end{subequations}  
\noindent where $\phi_{qp}(i',\tilde{\gamma})=(f_{q}(i',\tilde{\gamma})-f_{p}(i',\tilde{\gamma}))$. It should be noted that we have taken $\mathcal{J}_{ij}^{\gamma\gamma'}$ to be a Heisenberg coupling but off-diagonal terms can readily be considered and give rise to terms $\sim S_{\alpha qp}^{\tilde{\gamma}}S_{\mu pq}^{\tilde{\gamma}}S_{\nu kl}^{\gamma'}$, where $\mu\neq\nu$. By substituting Eqns. \cref{commutatora,commutatorb,commutatorc} into Eqn. \ref{eqofmot2} and performing a spatial Fourier transform one recovers an expression for the Green's function equation of motion
\begin{equation}
\begin{split}
&G_{\tilde{\gamma}\tilde{\gamma}'}^{\alpha\beta}(\mathbf{q},\omega)=g_{\tilde{\gamma}\tilde{\gamma}'}^{\alpha\beta}(\omega)\delta_{\tilde{\gamma}\tilde{\gamma}'}\\ &\qquad+\sum_{\gamma'}g_{\tilde{\gamma}\tilde{\gamma}}^{\alpha x}(\omega)\mathcal{J}_{\tilde{\gamma}\gamma'}(\mathbf{q})G_{\gamma'\tilde{\gamma}'}^{x\beta}(\mathbf{q},\omega)\\&\qquad+\sum_{\gamma'}g_{\tilde{\gamma}\tilde{\gamma}}^{\alpha y}(\omega)\mathcal{J}_{\tilde{\gamma}\gamma'}(\mathbf{q})G_{\gamma'\tilde{\gamma}'}^{y\beta}(\mathbf{q},\omega)  \\&\qquad+\sum_{\gamma'}g_{\tilde{\gamma}\tilde{\gamma}}^{\alpha z}(\omega)\mathcal{J}_{\tilde{\gamma}\gamma'}(\mathbf{q})G_{\gamma'\tilde{\gamma}'}^{z\beta}(\mathbf{q},\omega),
\end{split}
\label{fulleq}
\end{equation} 
\noindent with the single-ion Green's function given by 
\begin{equation}
g_{\tilde{\gamma}\tilde{\gamma}'}^{\alpha\beta}(\omega)=\sum_{qp}\frac{S^{\tilde{\gamma}}_{\alpha qp}S^{\tilde{\gamma}'}_{\beta pq}\phi_{qp}}{\omega-(\omega_{p}-\omega_{q})},
\label{SingleionGreensFunction}
\end{equation}
\noindent where we have used translational symmetry to drop the site index on $\phi_{qp}$. To describe magnon excitations, we sum over transitions to and from the ground state. For collinear systems the expression for the Green's function (Eqn. \ref{fulleq}) decouples into three matrix equations for the $xx$, $yy$ and $zz$ spin wave modes, with the dimension of the Green's function matrix given by the size of the crystallographic unit cell \cite{Lane21:104}. Eqn. \ref{fulleq} is general and can be used to treat antiferromagnetic systems by doubling the unit cell to account for the differing mean field on antialigned spins \cite{Lane21:104_2}. However, for noncollinear systems, the use of an enlarged supercell is not convenient, and in the case of incommensurate magnetic structures this is not possible. In the next section we preset a general method for treating any single $\mathbf{Q}$ magnetic structure. 

\subsection{Rotating frame formalism}   
The scheme presented in the previous section cannot treat general noncollinear magnetic structures since, in the lab frame, $(x,y,z)$, each unit cell has a different mean field Hamiltonian (and hence different $g_{\gamma\gamma}(\omega)$) up to the period of the magnetic supercell. This deficiency can be overcome by transforming to a reference frame that rotates with the magnetic structure~\cite{Haraldsen09:21}, $(\tilde{x},\tilde{y},\tilde{z})$. In this rotating frame, the magnetic moment at each site is orientated along the $\tilde{z}$ axis. The spin vector in the lab frame can be related to the rotating frame by the rotation

\begin{equation}
\mathbf{S}_{i\gamma}=R_{i\gamma}\mathbf{\tilde{S}}_{i\gamma}
\nonumber
\end{equation}    

\noindent where $\mathbf{\tilde{S}}_{i\gamma}$ are the spin operators in the rotating frame. The rotation can be broken into two parts, the rotation of the spins within the unit cell onto a common coordinate system for the unit cell and a rotation of each unit cell onto a common rotating frame coordinate system, $R_{i\gamma}\to R_{i}R_{\gamma}$. In order to relate $R_{i}$ to the magnetic ordering wavevector, $\mathbf{Q}$, and spin rotation plane, $\mathbf{n}$, we make use of the Rodrigues formula

\begin{subequations}
\begin{align}
\label{Rodrigues}
R_{i}=e^{i\mathbf{Q}\cdot \mathbf{r}_{i}}T+e^{-i\mathbf{Q}\cdot \mathbf{r}_{i}}T^{*}+\mathbf{n}\mathbf{n}^{T}\\
T=\frac{1}{2}\left(\mathbf{1}-\mathbf{n}\mathbf{n}^{T}-i[\mathbf{n}]_{\times}\right).\label{Rodriguesb}
\end{align} 
\end{subequations}

\noindent The matrix elements of the skew symmetric matrix can be conveniently written using the Levi-Civita symbol in Einstein notation, ${\left[[\mathbf{n}]_{\times}\right]^{i}}_{j}={\epsilon_{i}}^{jk}n_{k}$. 

In the rotating frame, the inter-site exchange Hamiltonian becomes
\begin{equation}
\begin{split}
\mathcal{H}_{int}=&\frac{1}{2}\sum_{ij}^{\gamma\gamma'}\mathbf{S}_{i\gamma}^{T}\mathcal{J}_{ij}^{\gamma\gamma'}\mathbf{S}_{j\gamma'}\\
 =&\frac{1}{2}\sum_{ij}^{\gamma\gamma'}\mathbf{\tilde{S}}_{i\gamma}^{T}R^{T}_{\gamma}R^{T}_{i}\mathcal{J}_{ij}^{\gamma\gamma'}R_{j}R_{\gamma'}\mathbf{\tilde{S}}_{j\gamma'}.
\end{split}
\nonumber
\end{equation}
\noindent To proceed with the calculation, we write $\mathbf{S}_{i\gamma}$ using the basis vectors of the space formed by the tensor product of the sublattice space and $\mathbb{R}_{3}$, $\mathbf{V}_{3N}=\mathbf{V}_{N}\otimes \mathbb{R}_{3}$, where $N$ is the number of sites in the unit cell. Though competing Heisenberg exchange can give rise to noncollinear order, many noncollinear magnetic systems in nature arise due to more complicated exchange terms including Dzyaloshinskii-Moriya and other off-diagonal couplings. These terms can be motivated on symmetry grounds \cite{Dzyaloshinsky58:4} and arise due to third order processes, involving exchange between excited spin-orbit levels \cite{Moriya60:120,Yosida:book}. These can be readily incorporated into this model by defining the exchange matrix in the full $3N\times3N$-dimensional space $\mathbf{V}_{3N}$ as 

\begin{equation}
\underline{\underline{\mathcal{J}}}^{\gamma\gamma'}=\begin{pmatrix}
J^{11}_{xx}&J^{11}_{xy}&J^{11}_{xz}&J^{12}_{xx}\\
J^{11}_{yx}&J^{11}_{yy}&J^{11}_{yz}&J^{12}_{yx}& \hdots\\
J^{11}_{zx}&J^{11}_{zy}&J^{11}_{zz}&J^{12}_{zx}\\
J^{21}_{xx}&J^{21}_{xy}&J^{21}_{xz}&\ddots\\
&\vdots\\
&&&&&J^{NN}_{xx}&J^{NN}_{xy}&J^{NN}_{xz}\\
&&&&&J^{NN}_{yx}&J^{NN}_{yy}&J^{NN}_{yz}\\
&&&&&J^{NN}_{zx}&J^{NN}_{zy}&J^{NN}_{zz}
\end{pmatrix}
\nonumber
\end{equation} 

\noindent forming a $3N\times 3N$ matrix. Note that for Heisenberg coupling, only the diagonal elements of each $3\times 3$ block are nonzero. In order that the rotation $R_{i\gamma}$ acts only within $\mathbb{R}_{3}$, we project into $\mathbf{V}_{3N}$, so that $T_{3N}=\mathbb{I}_{3}\otimes T$, where $\mathbb{I}_{3}$ is the $3 \times 3$ identity matrix and $T$ is defined by Eqn. \ref{Rodriguesb}. Since the rotation matrices are unitary, $R_{i}^{T}R_{j}=R_{ij}$, and the corresponding exponential factors from the Rodrigues formula (Eqn. \ref{Rodrigues}) can be absorbed into the definition of the Fourier transform of the exchange interaction. Expressed in the $3N\times 3N$ product space, the full inter-site  Hamiltonian can then be written

\begin{subequations}
\begin{gather}
\begin{split}
&\mathcal{H}_{int}=\frac{1}{2}\sum_{\mathbf{q}}\vec{\mathbf{\tilde{S}}}_{\mathbf{q}}^{T}\Big\{X'\Big[\underline{\underline{\mathcal{J}}}(\mathbf{q}+\mathbf{Q})T_{3N}\\&+\underline{\underline{\mathcal{J}}}(\mathbf{q}-\mathbf{Q})T_{3N}^{*}+\underline{\underline{\mathcal{J}}}(\mathbf{q})(\mathbb{I}_{3}\otimes\mathbf{n}\textbf{n}^{T} )\Big]X\Big\}\vec{\mathbf{\tilde{S}}}_{-\mathbf{q}}
\end{split}\\
\left[\underline{\underline{\mathcal{J}}}(\mathbf{q})\right]_{\gamma\gamma'}=\sum_{ij}\underline{\underline{\mathcal{J}}}_{ij}^{\gamma\gamma'}e^{-i\mathbf{q}\cdot(\mathbf{r}_{i}-\mathbf{r}_{j})}\label{Jmat:eq}\\
X=\mathrm{diag}\left(R_{1},R_{2},...,R_{N}\right)\\
X'=\mathrm{diag}\left(R^{T}_{1},R^{T}_{2},...,R^{T}_{N}\right)
\end{gather}
\end{subequations}

\noindent where $\vec{\mathbf{\tilde{S}}}_{\mathbf{q}}^{T}=(\tilde{S}^{x}_{1}(\mathbf{q}),\tilde{S}^{y}_{1}(\mathbf{q}),...,\tilde{S}^{z}_{N}(\mathbf{q}))$. The contents of the braces, $\{\}$, can be identified as a rotated exchange parameter, $\underline{\underline{\mathcal{\tilde{J}}}}(\mathbf{q})$, defined such that, $\mathcal{H}_{int}=\frac{1}{2}\sum_{\mathbf{q}}\vec{\mathbf{\tilde{S}}}_{\mathbf{q}}^{T}\underline{\underline{\mathcal{\tilde{J}}}}(\mathbf{q})\vec{\mathbf{\tilde{S}}}_{\mathbf{-q}}$. Even for Heisenberg coupling, this is no longer diagonal in $\mathbb{R}_{3}$ and contains terms that couple orthogonal modes. 

Note that we have performed the summation in Eqn. \ref{Jmat:eq} over the unit cell rather than over all sites as is required in the definition of the dynamical structure factor. This allows us to absorb the exponential factors from the Rodrigues formula (Eqn. \ref{Rodrigues}) into the definition of $\underline{\underline{\mathcal{J}}}$. The effect of summing over the unit cell is to create interference between the ions in the unit cell and can thus be regarded as a type of form factor.

In this rotated coordinate system, the calculation can be performed in a manner similar that outlined in the previous section, except in our new coordinate frame the coupling is not in general a diagonal Heisenberg coupling. The Green's function in the rotating frame can be written down by inspection of Eq. \ref{fulleq}, noting that in our new rotating frame $\underline{\underline{\mathcal{\tilde{J}}}}$ can couple orthogonal modes, hence

\begin{equation}
\begin{split}
&\tilde{G}_{\tilde{\gamma}\tilde{\gamma}'}^{\alpha\beta}(\mathbf{q},\omega)=g_{\tilde{\gamma}\tilde{\gamma}'}^{\alpha\beta}(\omega)\delta_{\tilde{\gamma}\tilde{\gamma}'}\\ &\qquad+\sum_{\gamma'}^{\mu\nu}g_{\tilde{\gamma}\tilde{\gamma}}^{\alpha \mu}(\omega)\tilde{\mathcal{J}}^{\mu\nu}_{\tilde{\gamma}\gamma'}(\mathbf{q})\tilde{G}_{\gamma'\tilde{\gamma}'}^{\nu\beta}(\mathbf{q},\omega)
\end{split}
\label{fulleqRF}
\nonumber
\end{equation} 

\noindent where $\tilde{G}(\mathbf{q},\omega)$ is the Green's function in the rotation frame This can be solved as a matrix equation in a manner similar to that described in Ref. \onlinecite{Lane21:104}. All that remains is to rotate back into the lab frame
\begin{equation}
    \begin{split}
    \underline{\underline{G}}(\mathbf{q},\omega)=D_{\mathbf{q}}(\mathbb{I}_{3}\otimes\mathbf{n}\textbf{n}^{T})X\underline{\underline{\tilde{G}}}(\mathbf{q},\omega)X^{\prime}(\mathbb{I}_{3}\otimes\mathbf{n}\textbf{n}^{T})D_{-\mathbf{q}}\\+D_{\mathbf{q}}T_{3N}^{*}X\underline{\underline{\tilde{G}}}(\mathbf{q}+\mathbf{Q},\omega)X^{\prime}T_{3N}^{\prime}D_{-\mathbf{q}}\\+D_{\mathbf{q}}T_{3N}X\underline{\underline{\tilde{G}}}(\mathbf{q}-\mathbf{Q},\omega)X^{\prime}T_{3N}^{*\prime}D_{\mathbf{-q}}
    \end{split}
    \nonumber
\end{equation}
\noindent where $T_{3N}^{\prime}=(\mathbb{I}_{3}\otimes T^{T})$ and $T_{3N}^{*\prime}=(\mathbb{I}_{3}\otimes T^{\dagger})$ and the translational invariance of the correlation function has been used. The matrix $D_{\mathbf{q}}=\delta_{\gamma\gamma'}e^{i\mathbf{q}\cdot\delta_{\gamma}}\otimes \mathbb{I}_{3}$ accounts for the interference between ions in the unit cell. If the ordering wavevector is $\mathbf{Q}=(0,0,0)$, we can perform the sum over all ions in the Fourier transform of the exchange interaction, $\underline{\underline{\mathcal{J}}}(\mathbf{q})$, in which case the Green's function in the lab frame is simply $\underline{\underline{G}}(\mathbf{q},\omega)=X\underline{\underline{\tilde{G}}}(\mathbf{q},\omega)X^{\prime}$.

\section{Application to rubidium iron fluoride}

We now turn our attention to the low energy magnetic fluctuations in the noncollinear antiferromagnet RbFe$^{2+}$Fe$^{3+}$F$_{6}$. The crystal structure of RbFe$^{2+}$Fe$^{3+}$F$_{6}$ is in the $Pnma$ space group (No. 62), with lattice parameters $a =6.9663(4)$,
 $b =7.4390(5)$ and $c =10.1216(6)$ \AA~ at T = 4 K~\cite{Kim11:3}. The charge order originates from the differing valence on the two Fe sites, with one site occupied by an Fe$^{2+}$ ion and the other by an Fe$^{3+}$ ion, (henceforth referred to site A and B respectively). Consequently, the two ions have different single-ion ground states, the former having an orbital degree of freedom, with $S=2, L=2$ and the latter being an orbital singlet, $S=5/2, L=0$. As a result, whilst a projection onto a spin-only Hamiltonian is well-justified for the Fe$^{3+}$ ions, the same is not necessarily true of the Fe$^{2+}$ ions, where evidence of the influence of orbital physics in the correlated magnetic behavior has already been reported \cite{Bai21:17}. 
 
 \begin{figure*}[h!t]
    \begin{center}
    \includegraphics[width=\linewidth]{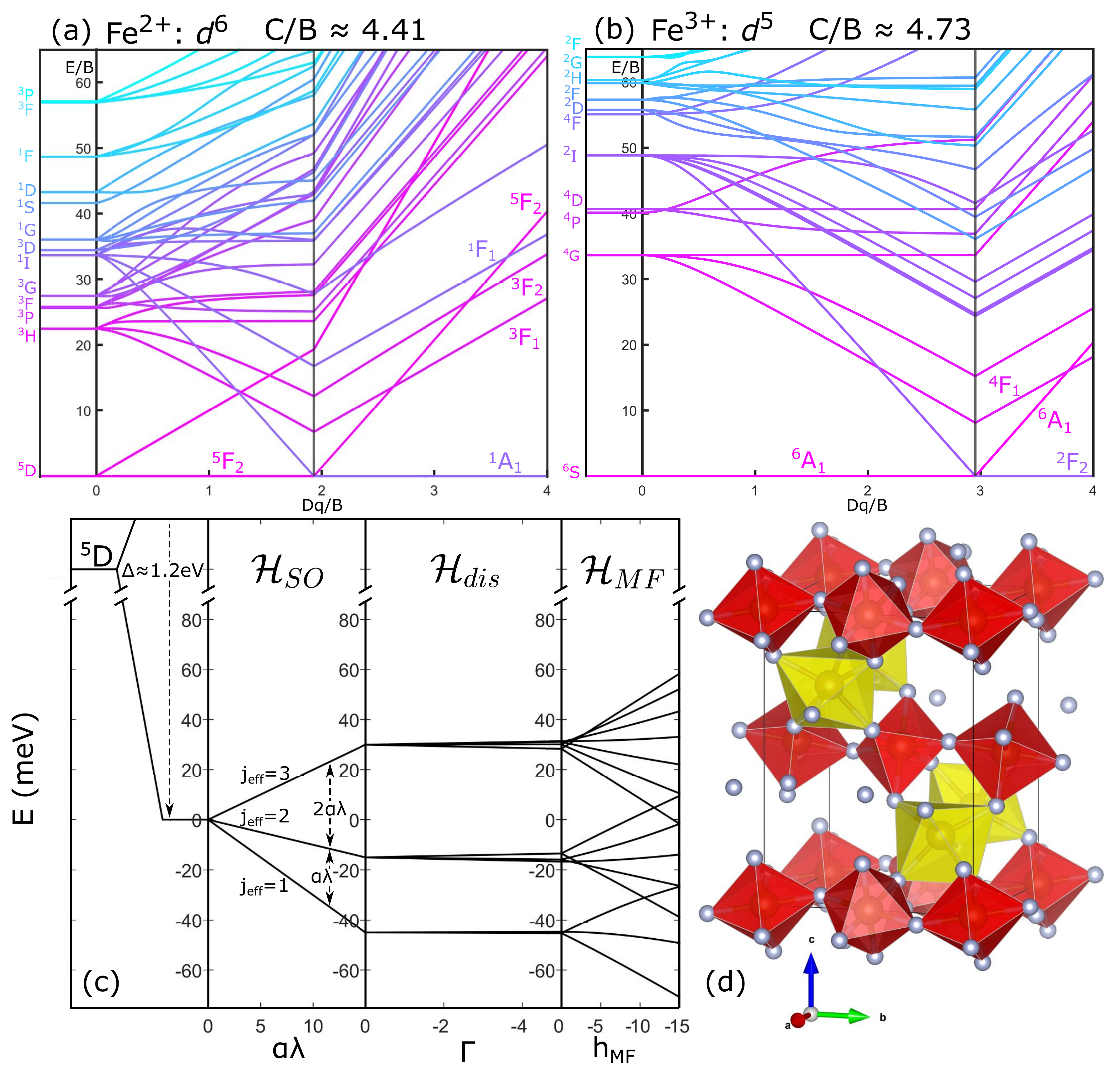}
    \end{center}
    \caption{$(a)$ Tanabe-Sugano diagram for a $d^{6}$ ion with Dq characterizing the strength of the octahedral crystal field and B and C the Racah parameters~\cite{Abragam:book}. Dq/B $\approx 1.1$ \cite{Abragam:book,McClure:book}, hence we cannot neglect orbital angular momentum and instead have a $(S=2, L=2)$ ground state. The Racah parameter, C $\approx$ 0.5 eV \cite{McClure:book}. $(b)$ Tanabe-Sugano diagram for a $d^{5}$ ion, showing the high spin to low spin transition at Dq/B $\approx3$. For Fe$^{3+}$ ions in an octahedral environment,  Dq/B$ \approx 1.6$ \cite{Abragam:book,McClure:book} motivating a spin only $(S=5/2, L=0)$ ground state. The Racah parameter C $\approx$ 0.6 eV~\cite{McClure:book}. $(c)$ single-ion energy levels for an Fe$^{2+}$ ion under the influence of spin-orbit coupling, crystallographic distortions and a molecular mean field term, following the intermediate crystal field splitting which gives rise to the $l=1$ ground state.  $(d)$ Crystal structure of RbFe$^{2+}$Fe$^{3+}$F$_{6}$, showing the octahedral FeF$_{6}$ coordination. Red octahedra surround Fe$^{3+}$ ions and yellow octahedra surround Fe$^{2+}$ ions. Figure created using VESTA~\cite{Momma11:44}.}
    \label{SingleionFig}
\end{figure*}
 
 The advantage of formulating the calculation in the manner described above is that one can explicitly treat the single-ion physics of the coupled magnetic ions, thus capturing the entangled nature of the spin and orbital degrees of freedom. Not only can inclusion of these single-ion terms change the energetics of the elementary excitations of the system, but terms such as spin-orbit coupling can lead to the non-conservation of $\hat{S}_{z}$ giving rise to correlated fluctuations of the spin amplitude in the form of longitudinal modes. Such modes are absent from conventional linear spin wave theory treatments. We now turn our attention to the single-ion physics of the Fe ions present in RbFe$^{2+}$Fe$^{3+}$F$_{6}$.

\subsection{Single-ion physics}

Whilst both Fe$^{2+}$ and Fe$^{3+}$ ions in the unit cell are surrounded by an octahedral environment of fluorine, the different sublattices are occupied by ions with a differing valence and local distorted environments. As a result, the ground state differs between sublattice A (Fe$^{2+}$) and sublattice B (Fe$^{3+}$). In this section (schematically outlined in Fig. \ref{SingleionFig}) we discuss the single-ion physics on both these sites which defines the eigenstates that we couple up using the random phase approximation discussed above.

\subsubsection{Sublattice A - Fe$^{2+}$ Single-ion Physics}

Sublattice A is occupied by Fe$^{2+}$ ions which are in the $3d^{6}$ configuration. Since the $3d$ ions experience an intermediate ligand field~\cite{McClure:book}, the single-ion ground state can be determined by the application of the Pauli exclusion principle and Hund's rules. Correspondingly, the ground state of the Fe$^{2+}$ ions is $^{5}D$ ($S=2, L=2$, or $|L=2, m_{L}; S=2, m_{S}\rangle$) (Fig. \ref{SingleionFig} $(a)$).

We first consider the strong crystalline electric field imposed on the Fe$^{2+}$ by the locally coordinated fluorine atoms, denoted as $\mathcal{H}_{CEF}$, on the orbital component and then discuss the effects of spin-orbit coupling below.  Using Stevens operators~\cite{Stevens52:65,Hutchings64:16}, for a $d^{6}$ ion in an octahedral crystal field this can be written as

\begin{equation}
\mathcal{H}_{CEF}=B_{4}(\mathcal{O}_{4}^{0}+5\mathcal{O}_{4}^{4}).
\nonumber
\end{equation} 

\noindent The fivefold degenerate $|L=2, m_{L}\rangle$ states are split into a ground state orbital triplet and an excited doublet. The crystal field splitting for $3d$ ions is on the order of $\sim$ 1 eV making this the largest single-ion energy scale. We note that simplistic point charge calculations enjoy limited success in the treatment of the $3d$ ions owing, in part, to the significant role played by covalency effects \cite{Avram:book}. Nonetheless, by measuring the crystal field splitting using optical spectra or RIXS, approximate values for the Stevens parameters can be extracted. The crystal field splitting in octahedrally-coordinated Fe$^{2+}$ was determined to be $10Dq \approx$ 1.2 eV \cite{Burns:book}, corresponding to $B_{4}\approx$ 10 meV.  A similar energy scale has been reported~\cite{Larson07:99,Haverkort07:99} in the oxides CoO~\cite{Cowley13:88} and NiO~\cite{Kim11:84}.   Writing the Stevens parameters in terms of the orbital angular momentum operators and using the notation for changing between $|L,m_{L} \rangle$ and the crystal field bases written in Refs. \onlinecite{Sarte20:102,Sarte19:100}, we can diagonalize the crystal field Hamiltonian

\begin{equation}
E_{CEF}=\mathcal{C}^{-1}\mathcal{H}_{CEF}\mathcal{C}=B_{4}\left(\begin{array}{ccc|cc}
\mathbf{-48}&\mathbf{0}&\mathbf{0}&0&0\\
\mathbf{0}&\mathbf{-48}&\mathbf{0}&0&0\\
\mathbf{0}&\mathbf{0}&\mathbf{-48}&0&0\\
\hline
0&0&0&\mathbf{72}&\mathbf{0}\\
0&0&0&\mathbf{0}&\mathbf{72}
\end{array}\right)
\nonumber
\end{equation}

\noindent and verify that the ground state orbital triplet is well separated from the excited orbital doublet. This is verified by the Tanabe-Sugano diagram for Fe$^{2+}$ reproduced in Fig. \ref{SingleionFig} $(a)$ with Dq/B $\approx$ 1.1 \cite{Abragam:book,McClure:book} called the weak-intermediate crystal field limit.  This orbital ground state is referred to as $^5D$ in Fig. \ref{SingleionFig} $(c)$.

Given the ground state is an orbital triplet, we are justified in projecting our single-ion Hamiltonian into an effective $l=1$ manifold.  This transformation carries a projection factor $\mathbf{L}=\alpha\mathbf{l}$~\cite{Abragam:book}  which can be read off the $l=1$ block of the $\hat{L}_{z}$ operator projected into the space spanned by the eigenvectors of $\mathcal{H}_{CEF}$

\begin{equation}
\mathcal{C}^{-1}\hat{L}_{z}\mathcal{C}=\left(\begin{array}{ccc|c|c}
\mathbf{-1}&\mathbf{0}&\mathbf{0}&0&0\\
\mathbf{0}&\mathbf{0}&\mathbf{0}&0&2\\
\mathbf{0}&\mathbf{0}&\mathbf{1}&0&0\\
\hline
0&0&0&\mathbf{0}&0\\
\hline
0&2&0&0&\mathbf{0}
\end{array}\right)
\nonumber
\end{equation}
\noindent thus $\alpha=-1$. Similar transformations for $\hat{L}_{x,y}$ show that this orbital triplet follows the correct commutator and Lie algebra for angular moment operators with $l=1$.  We note that this is not guaranteed based on degeneracy alone as discussed in Ref. \onlinecite{Pasztorova19:99} for the case of Ce$^{3+}$ in CeRhSi$_{3}$ in a comparatively anisotropic crystal field.

Having defined the orbital ground state, we define the new basis states to include spin as $|l=1, m_{l}, S=2, m_{s}\rangle$.  The next term to be considered is the spin-orbit interaction, denoted as $\mathcal{H}_{SO}$ in Fig. \ref{SingleionFig} $(c)$ acting on the projected orbital triplet with spin $S=2$ ($|l=1, m_{l};S=2, m_{s}\rangle$) (referred to as $^5D$ in Fig. \ref{SingleionFig} $(c)$). In terms of the projected orbital angular momentum, this can be written as
\begin{equation}
\mathcal{H}_{SO}=\lambda\mathbf{L}\cdot\mathbf{S}=\alpha\lambda\mathbf{l}\cdot\mathbf{S}
\nonumber
\end{equation} 

\noindent where $\lambda$ is the spin-orbit constant, which is negative for a greater-than-half-full outer shell \cite{Yosida:book}. For the free Fe$^{2+}$ ion $\alpha\lambda\approx$ 12.4 meV.~\cite{Abragam:book,McClure:book} This value is expected to be reduced due to the bonding with surrounding ligands, however this correction is expected to be small and is difficult to disentangle from the effects of Jahn-Teller distortions \cite{Abragam:book}, so we will neglect this correction from our analysis. The spin-orbit coupling splits the triply degenerate $l=1$ level into three $j_{eff}$ levels

\begin{equation}
\mathcal{C}^{-1}\mathcal{H}_{SO}\mathcal{C}= \alpha\lambda
\begin{pmatrix}
-3\mathbb{I}_{3}&\mathbf{0}&\mathbf{0}\\
\mathbf{0}&-\mathbb{I}_{5}&\mathbf{0}\\
\mathbf{0}&\mathbf{0}&2\mathbb{I}_{7}
\end{pmatrix}
\nonumber
\end{equation}

\noindent that follow the Land{\'e} interval rule. For a $3d^{6}$ ion, the ground state is the triply degenerate $j_{eff}=1$ level, with an excited quintet and septet. 

Further to the octahedral crystal field described earlier, the effect of distortions away from the perfect octahedral coordination must be considered. The octahedron surrounding the Fe$^{2+}$ is subtly compressed with four Fe-F bonds of length $\approx$ 2.1\AA~ and two of length $\approx$ 2.0 \AA. A tetragonal distortion of this kind can be described in terms of the Stevens operator

\begin{equation}
\label{tetragonaldistortion}
\mathcal{H}_{dis}=B_{2}^{0}\mathcal{O}_{2}^{0}=\Gamma \left(\hat{l}_{z}^{2}-\frac{2}{3}\right).
\end{equation}  

\begin{figure}[h]
    \centering
    \includegraphics[ trim={10.5mm 0mm 10.5mm 0mm},width=\linewidth]{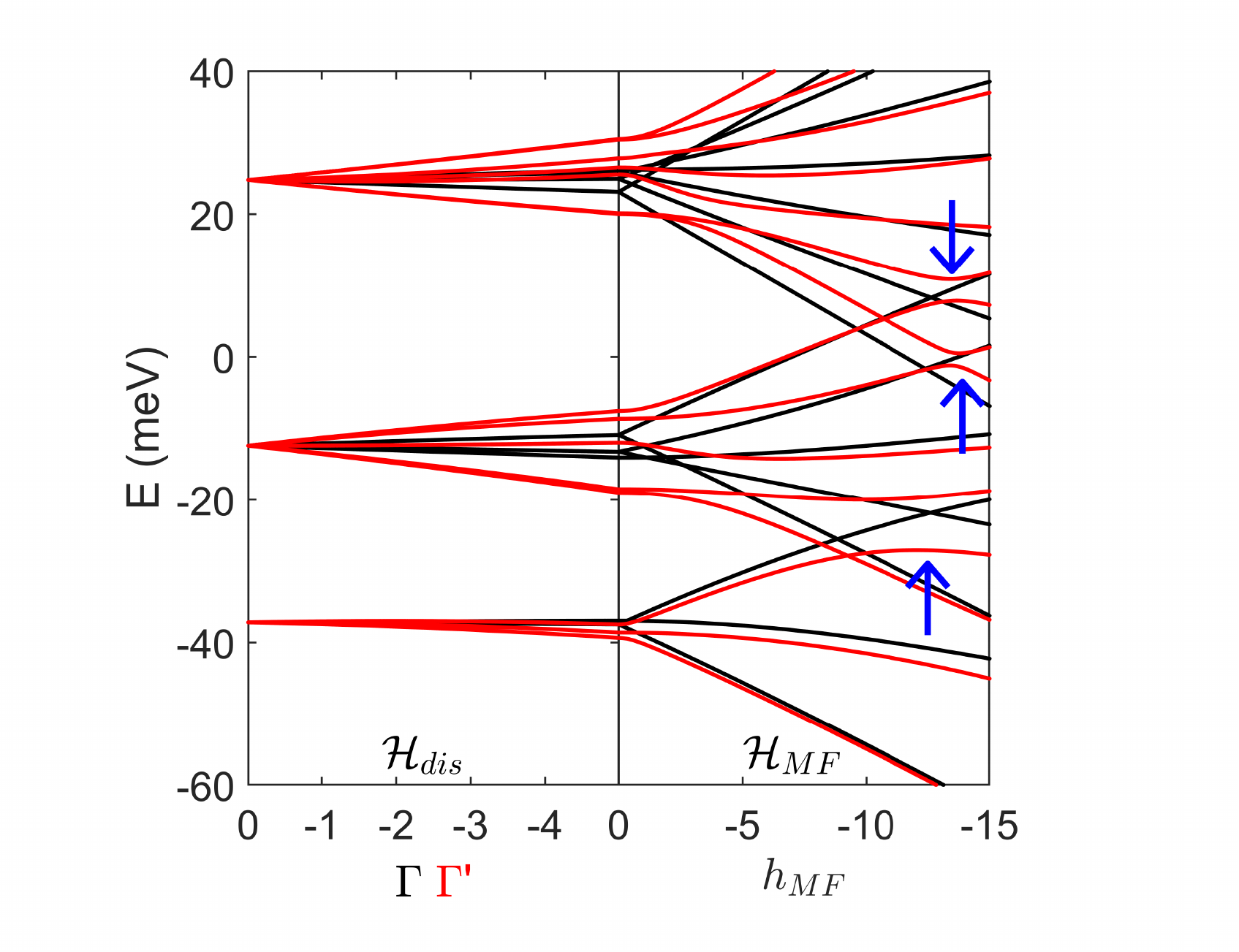}
    \caption{Energy diagram for Fe$^{2+}$ with spin-orbit coupling and crystallographic distortions. The black line represents a tetragonal distortion (Eqn. \ref{tetragonaldistortion}). The red lines indicate a distortion of the type described by Eqn. \ref{distortion22}. A tetragonal distortion gives rise to an orbital doublet. The asymmetric distortion give rise to splitting of the orbital triplet. Other qualitative differences can be seen, for example there are avoided crossings at around 12.5 meV, 13.5 meV and 13.9 meV for the asymmetric distortion (indicated by the blue arrows). The level repulsion at 12.5 meV is between the $j_{eff}=1$ and $j_{eff}=2$ multiplets.}
    \label{Distortionfig}
\end{figure}

\noindent The parameter $\Gamma$ is negative for an octahedral compression. This term breaks the triplet orbital degeneracy, leading to a doublet ground state with an excited singlet. In addition to this distortion, the octahedra are twisted in a manner which destroys the fourfold axial symmetry. Since the point group of the octahedron surrounding the Fe$^{2+}$ ion is the low symmetry $C_{1h}=C_{s}$ group, in principle, other terms of the form 
\begin{equation}
    \mathcal{H}_{CEF}=\sum_{kq}B_{k}^{q}\mathcal{O}_{k}^{q}
\end{equation}

\noindent are possible. The number of terms that must be considered can be reduced by a number of symmetry and physical considerations. The first is that since the Stevens operators depend on the tesseral harmonics, only terms for which the tesseral harmonics respect the point symmetry of the local crystal environment $(\mathcal{C}_{1h})$ are nonzero~\cite{GorllerWalrand:book}. The next consideration is that terms with $k>4$ vanish in the $3d$ ions since the matrix elements of the crystal field Hamitonian depend on the product of two spherical harmonics $Y_{2}^{-m}(\mathbf{R})Y_{2}^{m}(\mathbf{R})$ (where $k=2$ since we have $d$ electrons). From the Clebsch-Gordon expansion of this product, we find that the terms with $k>4$ vanish \cite{Bauer:book}. The Stevens parameters are given by \cite{Bauer:book} 

\begin{subequations}
\begin{align}
    B_{k}^{q}&=-|e| p_{k}^{q}\langle r^{k} \rangle \gamma_{k}^{q}\Theta_{k}\\
    \gamma^{q}_{k}&=\frac{1}{2k+1}\int d^{3}\mathbf{R}\frac{\rho (\mathbf{R})Z_{k}^{q}(\mathbf{R})}{\epsilon_{0}R^{k+1}}
    \label{gamma}
\end{align}
\nonumber
\end{subequations}

\noindent where $Z^{q}_{k}$ are the tesseral harmonics, with related numerical coefficients $p_{k}^{q}$, $\rho$ is the electrostatic charge density and $\Theta$ is a numerical factor originating from the conversion between polynomials and their operator equivalents \cite{Bauer:book}. For $k=2,4,6$, $\Theta_{k}$ are the well-known Stevens coefficients $\alpha_{J}$, $\beta_{J}$, $\gamma_{J}$ \cite{Stevens52:65}. The evaluation of the integral (Eqn. \ref{gamma}) is not a simple task. Practical calculations generally rely on vast simplifications such as a point-charge approximation which, as discussed previously, does not lead to quantitatively accurate predictions. It is therefore more appropriate to treat $B_{k}^{q}$ as experimentally determined parameters. Since the magnitude of $B^{q}_{k}$ scales as $\frac{1}{R^{k+1}}$, where $R$ is the distance from the central ion to the charged ligand, we can exclude the higher order terms since their effect will likely be small, we therefore exclude terms with $k>2$. Finally, the crystal field potential must satisfy time reversal symmetry~\cite{Tinkham:book}, hence we are left with one further possible distortion term   

\begin{equation}
\mathcal{H}^{\prime}_{dis}=\Gamma^{\prime}\left(l_{+}^{2}+l_{-}^{2}\right)\label{distortion22}
\end{equation}

\noindent where we have converted to operator equivalent terms and collected all factors into a single distortion parameter~\cite{Danielsen:book}. The effect of this term is to break the remaining degeneracy of the orbital doublet. Notice that the additional term has the same form as the perturbation in the widely-studied Lipkin model \cite{Lipkin65:62}, which exhibits an exceptional point and a transition from a phase with an avoided crossing to one with a degeneracy \cite{Heiss12:45}. In fact, this term gives rise to avoided crossings at $h_{MF}\approx 12.5 \mathrm{meV}$, $h_{MF}\approx 13.5 \mathrm{meV}$ and $h_{MF}\approx 13 .9\mathrm{meV}$ (Fig. \ref{Distortionfig}), suggesting the presence of an exceptional point in the complex plane of $(\Gamma^{\prime},h_{MF})$, close to the real axis \cite{Heiss05:38}. These three identified instances of level repulsion also indicate that the single-ion eigenfunctions are strongly mixed between the $j_{eff}=1$ and $j_{eff}=2$ and the $j_{eff}=2$ and $j_{eff}=3$ manifolds.

The final term that must be considered in the single-ion Hamiltonian is the molecular mean field. The exchange interaction between magnetic ions results in an effective Zeeman term from the single-ion perspective. In order that we expand around the correct single-ion ground state, a mean field decoupling must be performed $\mathbf{S}_{i}\to\langle \mathbf{S}_{i}\rangle+\delta \mathbf{S}_{i}$ to quantify the strength of this effective Zeeman field. As outlined in the discussions above, in general a Heisenberg model can be written as 
\begin{equation}
    \mathcal{H}=\frac{1}{2}\sum_{ij}^{\gamma\gamma'}\mathcal{J}_{ij}^{\gamma\gamma'}\mathbf{S}^{T}_{i\gamma}\cdot\mathbf{S}_{j\gamma'}.
    \nonumber
\end{equation}
\noindent We can perform a mean field decoupling and discard terms $\sim \mathcal{O}(\delta \mathbf{S}_{i})^{2}$. In the rotating frame, we have
\begin{equation}
\begin{split}
\mathcal{H}_{MF}=&\frac{1}{2}\sum_{ij}^{\gamma\gamma'}\Big([\mathbf{\tilde{S}}^{T}_{i\gamma}-\frac{1}{2}\langle \mathbf{\tilde{S}}_{i\gamma}^{T}\rangle] \underline{\underline{\tilde{\mathcal{J}}}}_{ij}^{\gamma\gamma'}\langle\mathbf{\tilde{S}}_{j\gamma'}\rangle\\
&\qquad +\langle\mathbf{\tilde{S}}^{T}_{i\gamma}\rangle \underline{\underline{\tilde{\mathcal{J}}}}_{ij}^{\gamma\gamma'}[\mathbf{\tilde{S}}_{j\gamma'}-\frac{1}{2}\langle \mathbf{\tilde{S}}_{j\gamma'}\rangle] \Big),
\end{split}
\nonumber
\end{equation}
\noindent where $\underline{\underline{\tilde{\mathcal{J}}}}_{ij}^{\gamma\gamma'}=R_{\gamma}^{T}R_{i}^{T}\mathcal{J}_{ij}^{\gamma\gamma'}R_{j}R_{\gamma'}$. Neglecting constant terms, we can simplify this expression considerably, using the Rodrigues rotation formula
\begin{equation}
\label{meanfieldeq}
\begin{split}
    \mathcal{H}_{MF}=& \frac{1}{2}\sum_{ij}^{\gamma\gamma'}\mathbf{\tilde{S}}^{T}_{i\gamma}\Big( \underline{\underline{\tilde{\mathcal{J}}}}_{ij}^{\gamma\gamma'}+\underline{\underline{\tilde{\mathcal{J}}}}_{ji}^{\gamma'\gamma}\Big)\langle\mathbf{\tilde{S}}_{j\gamma'}\rangle\\
    =& \sum_{i\gamma}\mathbf{\tilde{S}}^{T}_{i\gamma}\sum_{j\gamma'}\mathrm{Re}\left[\underline{\underline{\tilde{\mathcal{J}}}}^{\gamma\gamma'}(\mathbf{Q})\right]\langle\mathbf{\tilde{S}}_{j\gamma'}\rangle.
\end{split}
\nonumber
\end{equation}
\noindent In the rotating frame, the expectation value of the spin operators only have nonzero $z$-components. For the $3d$ ions, the inter-ion coupling is predominantly described by a spin-spin Heisenberg model, owing to the breaking of the ground state orbital degeneracy due to crystallographic distortions or spin-orbit coupling \cite{Kugel82:25}. This motivates a spin-only inter-ion interaction.

Collecting all of these single-ion terms together, we find the single-ion Hamiltonian on sublattice A, 
\begin{equation}
    \mathcal{H}^{A}_{1}=\mathcal{H}_{SO}+\mathcal{H}_{dis}++\mathcal{H}^{\prime}_{dis}+\mathcal{H}_{MF}.
    \nonumber
\end{equation}

The presence of $\mathcal{H}_{dis}$ and $\mathcal{H}_{SO}$ terms in the single-ion Hamiltonian results in the non-conservation of $\hat{S}_{z}$. Thus longitudinal transitions are allowed between different single-ion energy levels. Longitudinal modes are present in noncollinear magnets due to the loss of spin rotational symmetry about $\hat{z}$ \cite{Zhitomirsky13:85} and give rise to anharmonic scattering terms corresponding to coupling between transverse magnons and the two particle continuum. Systems with non-trivial single-ion physics offer an exciting opportunity for the observation of correlated amplitude fluctuations, since the fundamental excitonic spectrum includes a longitudinal component.  

The effect of spin-orbit transitions between different $j_{eff}$ levels has been observed in, for example,  $\alpha,\gamma$-CoV$_{2}$O$_{6}$ \cite{Wallington15:92}, $\alpha$-Co$_{3}$V$_{2}$O$_{8}$~\cite{Sarte18:98}, CoTiO$_{3}$ \cite{Yuan20:102}, Na$_{3}$Co$_{2}$SbO$_{6}$ or Na$_{2}$Co$_{2}$TeO$_{6}$~\cite{Songvilay20:102,Kim21:34}, and CoO \cite{Cowley13:88,Sarte18:98}. The spin-orbit splitting is typically on the order of $\approx 30$ meV in $3d$ ions and hence these spin-orbit excitons may be expected to be short-lived due to a large kinematically-allowed decay region. The intensity of such modes depends strongly on the single-ion physics and whilst these spin-orbit transitions have been observed in Co$^{2+}$ ions, they have not been observed in some other $3d$ ions such as V$^{3+}$ \cite{Lane21:104}.    

The propensity for longitudinal modes to decay can be overcome by moving these amplitude fluctuations out of the kinematically-allowed decay region. Therefore, the search for long-lived amplitude fluctuations at low energy may be fruitful. Amplitude fluctuations may be observed in other $3d$ ions where the excitonic modes originate not from the $j_{eff} \to j_{eff}$ transitions but from a smaller splitting due to $\mathcal{H}_{dis}$. The intensity of these transitions depends strongly on the nature of the distortion and the resulting single ion energy levels.

\begin{figure*}[ht]
    \begin{center}
    \includegraphics[width=\linewidth]{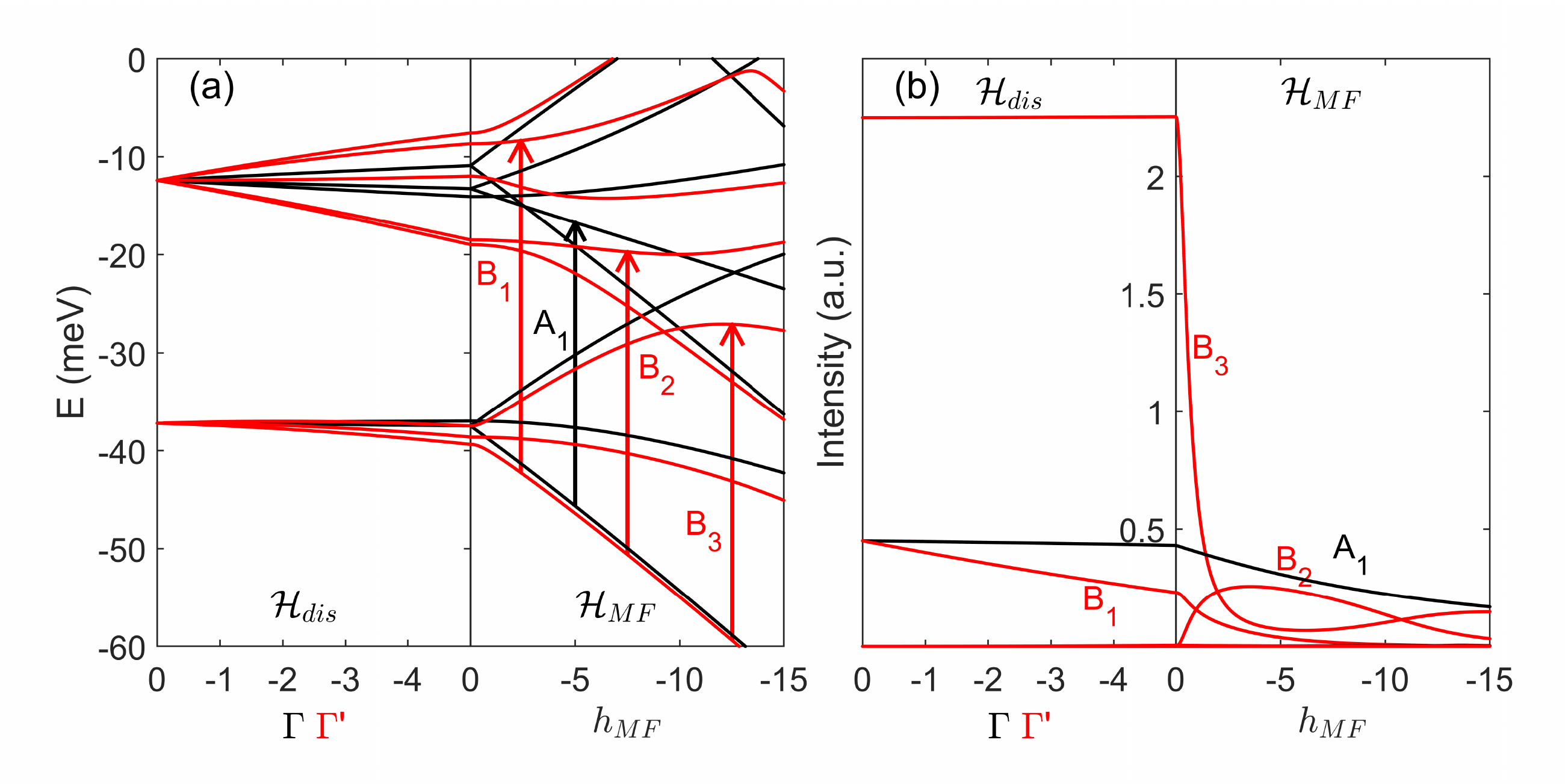}
    \end{center}
    \caption{$(a)$ Single-ion energy levels for Fe$^{2+}$ in a distorted octahedral environment. $(b)$ Longitudinal (or amplitude) transition amplitudes, $\mathcal{I}_{zz}= \lvert\bra{1}\hat{S}_{z}\ket{m} \rvert^{2}$, for the excitations out of the ground state. As $h_{MF}$ is increased the amplitude of the transitions A$_{1}$, B$_{1}$ and B$_{3}$ decreases. For a tetragonal/trigonal distortion ($\Gamma$), only the spin-orbit transition has non-negligible intensity. For the low symmetry distortion $(\Gamma')$, the high intensity transition B$_{3}$ corresponds to a transition within the ground state $j_{eff}=1$ multiplet. As the mean field increases, the inter-multiplet transition B$_{2}$ turns on and at large values of $h_{MF}$ the intensity of B$_{3}$ begins to increase again.}
    \label{LongitudinalTransitionsFig}
\end{figure*}

We now demonstrate that crystallographic distortions offer a mechanism for longitudinal excitons in $3d$ ions, but that a large molecular Zeeman field reduces the longitudinal transition amplitude for many of the transitions in Fe$^{2+}$ ions. The neutron scattering intensity is proportional to the transition amplitude $\mathcal{I}_{zz}= \lvert\bra{1}\hat{S}_{z}\ket{m} \rvert^{2}$. In Fig. \ref{LongitudinalTransitionsFig} we plot $\mathcal{I}_{zz}$ for the both the tetragonal and the asymmetric distortions introduced above. For both distortions, longitudinal transitions from the $j_{eff}=1$ to $j_{eff}=2$ have finite amplitude. For a tetragonal distortion, the transition A$_{1}$ is the sole longitudinal transition which carries non-negligible intensity. In the case of the asymmetric distortion, the B$_{1}$ transition loses intensity with increasing $h_{MF}$ and is overtaken by B$_{2}$. For the asymmetric distortion, an inter-multiplet mode, B$_{3}$, is also observed. As the mean field is increased, the intensity of most longitudinal modes decreases, although an increase in the intensity of B$_{2}$ is observed, along with an increase in B$_{3}$ at large values of $h_{MF}$, as the single-ion energy landscape changes. 

The longitudinal excitations described in this paper result from the non-conservation of $\hat{S}_{z}$. In other words, they correspond to amplitude fluctuations of the order parameter, in this case the spin operator. This motivates an analogy~\cite{Jain13:17,Souliou17:119} with the Higgs mechanism from particle physics, where amplitude fluctuations of the order parameter~\cite{Su20:102} in the presence of a gauge field give rise to the celebrated Higgs boson \cite{Higgs64:13}. The case here is somewhat different, owing to the lack of a coupling of the order parameter to a gauge field as in the Higgs mechanism. We shall therefore refer to these excitations as ``amplitude modes" to distinguish them both from the true gauge-field-coupled phenomena such as the Higgs boson \cite{Higgs64:13}, plasmons \cite{Anderson63:130}, the Meissner state in superconductors~\cite{Shimano20:11}, and from other longitudinal excitations whose origins are fundamentally different, such as spinons and multi-magnon continua \cite{Coldea03:68,Lake00:85}.  It is worth noting that these fluctuations can be observed with other experimental techniques with complementary selection rules to neutron scattering such as Raman~\cite{Souliou17:119}.

\subsubsection{Sublattice B - Fe$^{3+}$ single-ion physics}

In the case of a $3d^{5}$ ion in a perfectly octahedral environment, the ground state is an orbital singlet, $(S=5/2,L=0)$, hence we should only expect a mean molecular field contribution to the single-ion Hamiltonian. However, in many $3d^{5}$ systems, a spectral gap is measured, consistent with a single-ion anisotropy term \cite{Lane21:104_2,Calder19:99,deVries09:79}. This gap arises due to mixing of higher orbital energy levels into the ground state, facilitated by the cooperative effect of crystallographic distortions and spin-orbit coupling \cite{Watanabe57:18,Pryce50:80,Bleaney54:233}. We account for this phenomenologically in our model by adding a single-ion anisotropy term to the Fe$^{3+}$ spin Hamiltonian,   

\begin{subequations}
    \begin{align}
    \mathcal{H}^{B}_{1}=&\mathcal{H}_{MF}+\mathcal{H}_{anis}\\
    \mathcal{H}_{anis}=&\mu \tilde{S}_{z}^{2}.
    \end{align}
    \nonumber
\end{subequations}

\subsection{Spin Hamiltonian}
\begin{figure}[h!t]
    \begin{center}
    \includegraphics[width=\linewidth]{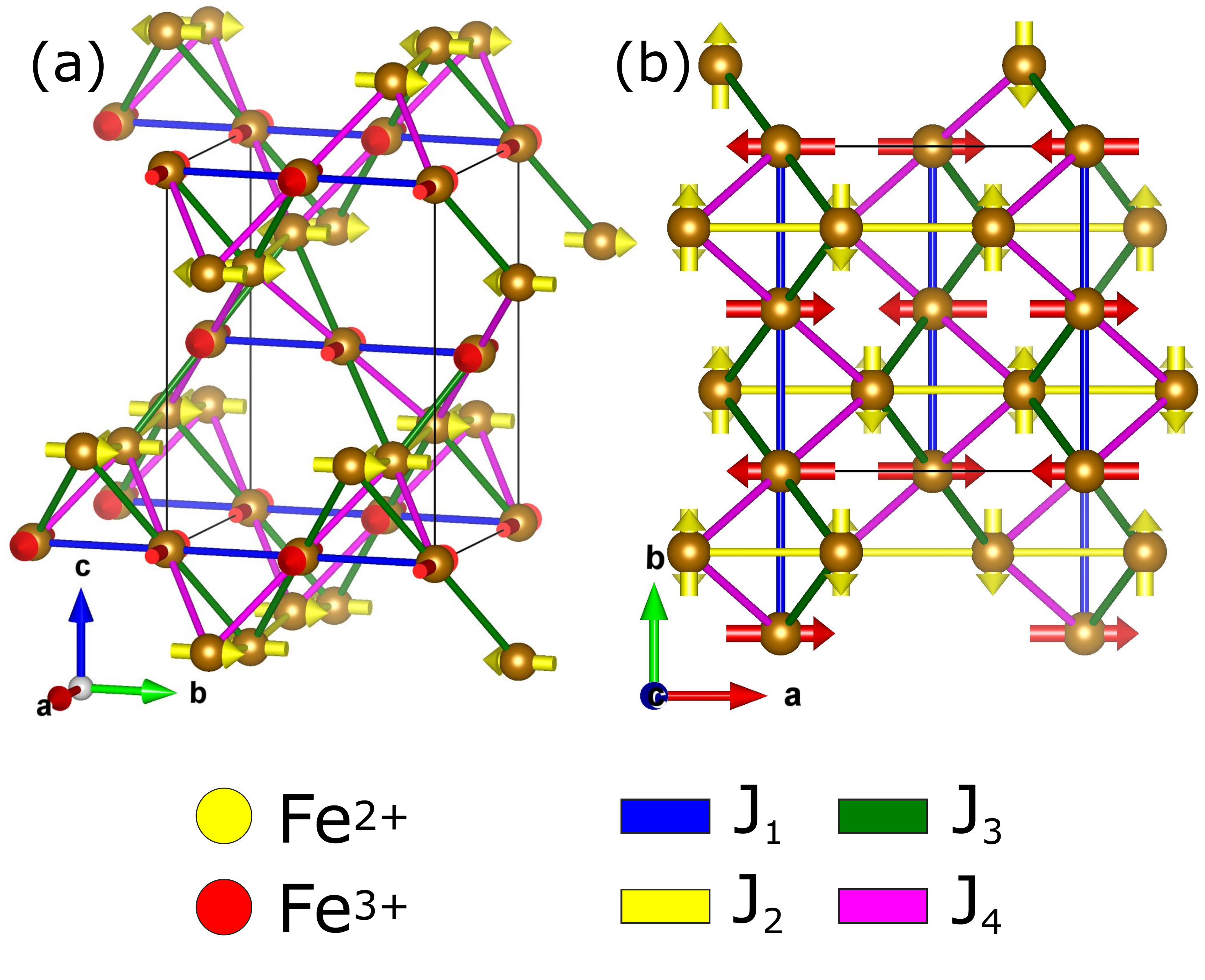}
    \end{center}
    \caption{$(a,b)$ Crystallographic structure of RbFe$^{2+}$Fe$^{3+}$F$_{6}$, displaying the Fe ions and Fe-Fe bonds. Yellow arrows indicate Fe$^{2+}$ ions and red arrows indicate Fe$^{3+}$ ions. The Fe$^{2+}$ ions form chains of spins along the $a$-axis and the Fe$^{3+}$ ions lie in chains along the $b$-axis. $J_{1}$ and $J_{2}$ are intra-chain bonds whilst $J_{3}$ and $J_{4}$ are inter-chain bonds. Figure created using VESTA~\cite{Momma11:44}.}
    \label{Exchanges_fig}
\end{figure}

We now turn our attention to the spin Hamiltonian that describes the interaction of ions on neighboring sites. The Fe ions in RbFe$^{2+}$Fe$^{3+}$F$_{6}$ form two interpenetrating chain networks running perpendicular to one another (Fig. \ref{Exchanges_fig}). The Fe$^{2+}$ ions lie on a chain parallel to $a$ with spins pointing along $\pm \hat{b}$, with Fe$^{3+}$ ions on a chain parallel to $b$ with spins along $\pm \hat{a}$. RbFe$^{2+}$Fe$^{3+}$F$_{6}$ can be described with a unit cell comprising eight spins (Table \ref{Unitcell}). 

\begin{table}[h]
\caption{\label{Unitcell} Definition of ions in the unit cell.}
\begin{ruledtabular}
\begin{tabular}{cccc}
Index & Sublattice & Valence & Position vector \\ 
\hline
1 &B & 3+ & (0,0,0)  \\ 
2 &B & 3+ & (0,0.5,0) \\ 
3 &B & 3+ & (0.5,0,0.5) \\ 
4 &B & 3+ & (0.5,0.5,0.5) \\ 
5 &A & 2+ & (0.1986,0.75,0.2698)  \\ 
6 &A & 2+ & (0.6986,0.75,0.2302) \\ 
7 &A & 2+ & (0.3014,0.25,0.7698) \\ 
8 &A & 2+ & (0.8014,0.25,0.7302) \\ 
\end{tabular}
\end{ruledtabular}
\end{table}   
We now consider a minimal model of nearest neighbor exchange for both inter- and intra-chain bonds. The exchange interactions are summarized below in Table \ref{Bonds}. The intra-chain bonds are confined to the the upper-left and lower-right blocks, with inter-chain bonds coupling sites $\{1$-$4\}$ with sites $\{5$-$8\}$.
\begin{table}[h]
\caption{\label{Bonds} Inter-site bonds considered in the minimal model. $J_{1}$ and $J_{2}$ represent intra-chain bonds with $J_{3}$ and $J_{4}$ coupling sites on different chains.    }
\begin{center}
\begin{tabularx}{0.475\textwidth} { 
  | >{\centering\arraybackslash}X
  | >{\centering\arraybackslash}X 
  | >{\centering\arraybackslash}X 
  | >{\centering\arraybackslash}X 
  | >{\centering\arraybackslash}X 
  | >{\centering\arraybackslash}X 
  | >{\centering\arraybackslash}X 
  | >{\centering\arraybackslash}X 
  | >{\centering\arraybackslash}X |}
 \hline
 &$\mathbf{1}$&$\mathbf{2}$&$\mathbf{3}$&$\mathbf{4}$&$\mathbf{5}$&$\mathbf{6}$&$\mathbf{7}$&$\mathbf{8}$ \\
 \hline
 $\mathbf{1}$&0&$J_{1}$&0&0&$J_{3}$&$J_{4}$&$J_{4}$&$J_{3}$  \\
 \hline
 $\mathbf{2}$&$J_{1}$&0&0&0&$J_{3}$&$J_{4}$&$J_{4}$&$J_{3}$ \\
 \hline
 $\mathbf{3}$&0&0&0&$J_{1}$&$J_{4}$&$J_{3}$&$J_{3}$&$J_{4}$ \\
\hline
$\mathbf{4}$&0&0&$J_{1}$&0&$J_{4}$&$J_{3}$&$J_{3}$&$J_{4}$ \\
\hline
$\mathbf{5}$&$J_{3}$&$J_{3}$&$J_{4}$&$J_{4}$&0&$J_{2}$&0&0 \\
\hline
$\mathbf{6}$&$J_{3}$&$J_{3}$&$J_{4}$&$J_{4}$&$J_{2}$&0&0&0 \\
\hline
$\mathbf{7}$&$J_{4}$&$J_{4}$&$J_{3}$&$J_{3}$&0&0&0&$J_{2}$ \\
\hline
$\mathbf{8}$&$J_{3}$&$J_{3}$&$J_{4}$&$J_{4}$&0&0&$J_{2}$&0 \\
\hline
\end{tabularx}
\end{center}
\end{table}
Following the approach outlined above, we now take the Fourier transform of the exchange interaction. In the rotating frame we need to calculate the matrix
\begin{equation}
\begin{split}
    \underline{\underline{\mathcal{\tilde{J}}}}(\mathbf{q})=&X'\Big[\underline{\underline{\mathcal{J}}}(\mathbf{q}+\mathbf{Q})T_{3N}+\underline{\underline{\mathcal{J}}}(\mathbf{q}-\mathbf{Q})T_{3N}^{*}\\&\qquad +\underline{\underline{\mathcal{J}}}(\mathbf{q})(\mathbb{I}_{3}\otimes\mathbf{n}\textbf{n}^{T})\Big]X.
\end{split}
\nonumber
\end{equation}
\noindent Since the propagation vector, $\mathbf{Q}=(0,0,0)$, we need not perform the rotation of each unit cell and instead have $\underline{\underline{\mathcal{\tilde{J}}}}(\mathbf{q})=X'\underline{\underline{\mathcal{J}}}(\mathbf{q})X$, where we have summed over all spins. The matrices, $X$ and $X'$ describe the matrices which rotate the spins in the unit cell onto a common axis. Since, in the lab frame, the spins lie in the $a$-$b$ plane, we can define a rotation matrix
\begin{equation}
    U(\theta)=
    \begin{pmatrix}
    0&\mathrm{sin}\theta&\mathrm{cos}\theta\\
    0&-\mathrm{cos}\theta&\mathrm{sin}\theta\\
    1&0&0
    \end{pmatrix}
    \nonumber
\end{equation}
\noindent that rotates spins by angle $\theta$ in the $a$-$b$ plane. In terms of this rotation matrix, we have
\begin{equation}
    X=\begin{pmatrix}
    U_{-a}&0&0&0&0&0&0&0\\
    0&U_{a}&0&0&0&0&0&0\\
    0&0&U_{a}&0&0&0&0&0\\
    0&0&0&U_{-a}&0&0&0&0\\
    0&0&0&0&U_{b}&0&0&0\\
    0&0&0&0&0&U_{-b}&0&0\\
    0&0&0&0&0&0&U_{-b}&0\\
    0&0&0&0&0&0&0&U_{b}
    \end{pmatrix},
    \nonumber
\end{equation}
\noindent where $U_{a}=U(0)$, $U_{-a}=U(\pi)$ and $U_{\pm b}=U(\pm \frac{\pi}{2})$, such that 
\begin{subequations}
    \begin{align}
        U_{a}
        \begin{pmatrix}
        0\\
        0\\
        1
        \end{pmatrix}&=
                \begin{pmatrix}
        1\\
        0\\
        0
        \end{pmatrix}\\
                U_{-a}
        \begin{pmatrix}
        0\\
        0\\
        1
        \end{pmatrix}&=
                \begin{pmatrix}
        -1\\
        0\\
        0
        \end{pmatrix}\\
                        U_{b}
        \begin{pmatrix}
        0\\
        0\\
        1
        \end{pmatrix}&=
                \begin{pmatrix}
        0\\
        1\\
        0
        \end{pmatrix}\\
                                U_{-b}
        \begin{pmatrix}
        0\\
        0\\
        1
        \end{pmatrix}&=
                \begin{pmatrix}
        0\\
        -1\\
        0
        \end{pmatrix}.
    \end{align}
    \nonumber
\end{subequations}
\noindent Using these rotation matrices, we can write down the molecular mean field Hamiltonian for each site. In this minimal model the mean field is the same for all spins on each sublattice,  
\begin{subequations}
\begin{align} 
   \mathcal{H}_{MF}=&\sum_{i\gamma}h_{MF}(i,\gamma)\tilde{S}^{z}_{i\gamma} \\
   h_{MF}(i,\gamma \in A)=&-2J_{2}\langle S_{A}\rangle=-4J_{2}\\
   h_{MF}(i,\gamma \in B)=&-2J_{1}\langle S_{B}\rangle=-5J_{1}.
\end{align}   
\nonumber
\end{subequations}
\noindent The molecular mean field does not depend on the inter-chain bonds since the spins on sublattice A are perpendicular to sublattice B.  

The matrix $\underline{\underline{\mathcal{J}}}(\mathbf{q})=\sum_{ij}\underline{\underline{\mathcal{J}}}_{ij}^{\gamma\gamma'}e^{i\mathbf{q}\cdot (\mathbf{r}_{i\gamma}-\mathbf{r}_{j\gamma'})}$ can be constructed from Tables \ref{Unitcell} and \ref{Bonds}, and is written out explicitly in Appendix \ref{AppendixA}. The exchange matrix in the lab frame contains only diagonal elements but on transforming to the rotating frame acquires components that couple $y$ and $z$ components of spins on different sublattices.
\section{Dynamical structure factor calculations}

\subsection{Parameter choice}

\begin{figure*}[h!t]
    \begin{center}
    \includegraphics[trim = {12.5mm 17.5mm 5mm 5mm},width=\linewidth]{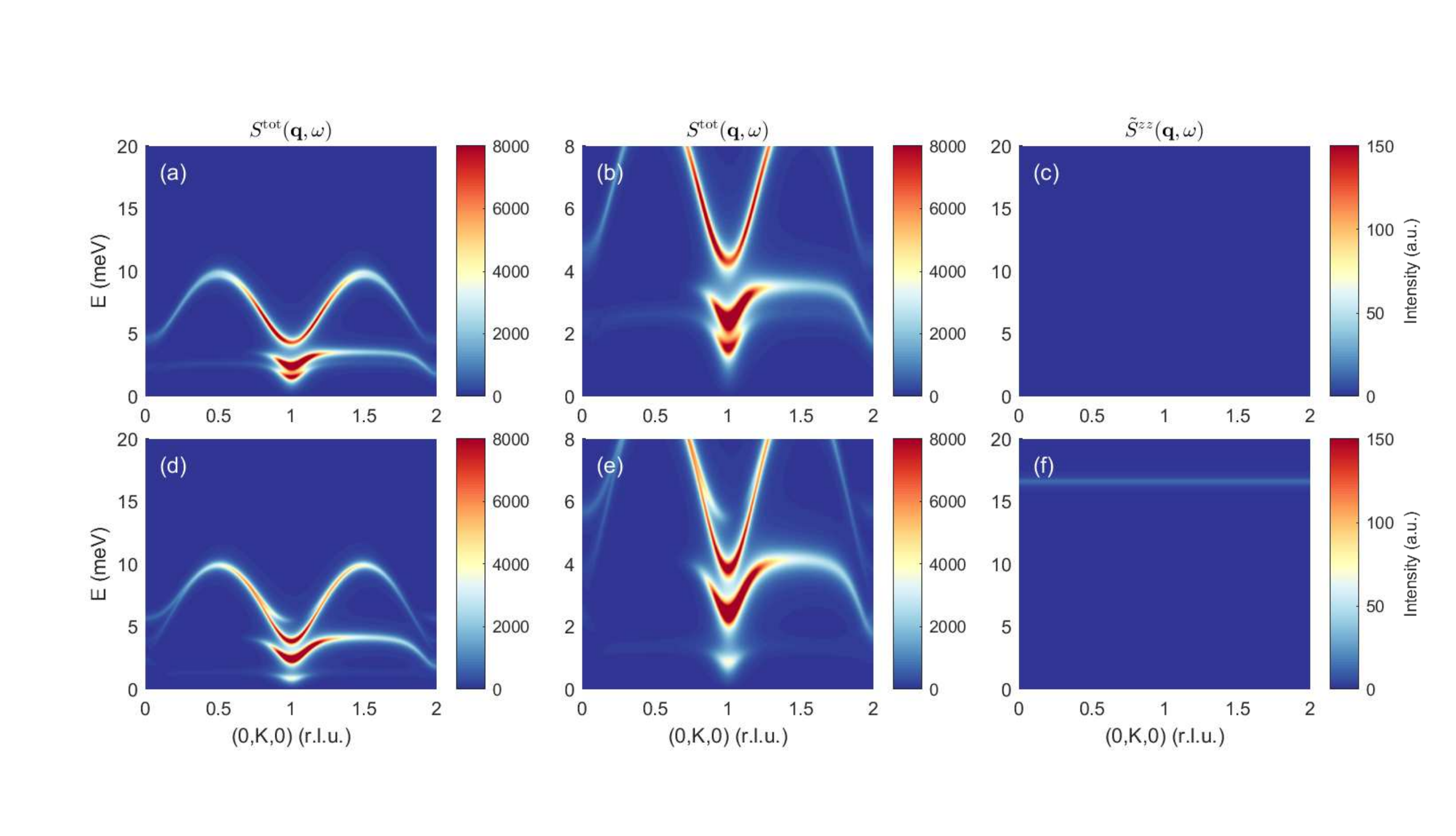}
    \end{center}
    \caption{Dynamical structure factor calculation for RbFe$^{2+}$Fe$^{3+}$F$_{6}$. In the right hand column is the longitudinal component in the rotating frame which contains the contribution from amplitude fluctuations. Panels $(a$-$c)$ show the calculated response with exchange parameters taken from Ref. \cite{Songvilay18:121}, along with a tetragonal compression, $\Gamma=-1.5$ meV. Panels $(d$-$f)$ show the corresponding calculation with an asymmetric distortion $\Gamma^{\prime}=-1.5$ meV.}
    \label{Realistic_fig}
\end{figure*}

We now use the rotating frame Green's function formalism to calculate the dynamical structure factor of RbFe$^{2+}$Fe$^{3+}$F$_{6}$. Samples of RbFe$^{2+}$Fe$^{3+}$F$_{6}$ produced using hydrothermal growth techniques are typically small rod-like crystals, with the long-axis coinciding with the crystallographic $b$-axis \cite{Songvilay18:121,Kim11:3}. Neutron scattering experiments thus necessitate the coalignment of many single crystals and a broad integration of spectral weight along directions perpendicular to the scattering wavevector, offering sensitivity to fluctuations along all three directions. Consequently, we drop the form factor and polarization factor in the structure factor and sum over all components of the partial dynamical structure factor, $S^{\mathrm{tot}}(\mathbf{q},\omega)=\sum_{\alpha\beta}S^{\mathrm{\alpha\beta}}(\mathbf{q},\omega)$. The parameters of the model are summarized in Table \ref{Parameter:Table}. Exchange parameters $J_{1}$-$J_{4}$ are taken from Ref. \onlinecite{Songvilay18:121}, along with the phenomenological anisotropy parameter $\mu$. The value of $\lambda$ was chosen in accordance with perturbative calculations and paramagnetic resonance of Fe$^{2+}$ in MgO \cite{Low60:118,Low60:118_2}. The distortion parameter, $\Gamma$, is chosen to be small, on the order of meV ($\Gamma=-1.5$ meV), consistent in scale with the parameter extracted from fits to neutron data in Co$^{2+}$ \cite{Cowley73:6,Sarte19:100} and V$^{3+}$ \cite{Lane21:104} compounds.
\begin{table}[h!]
\centering
\begin{tabular}{cc}
    \hline \hline
    Parameter & Value (meV)\\ \hline
    $J_{1}$ & 1.9 \\
    $J_{2}$ & 1.4 \\
    $J_{3}$ & 1.4 \\
    $J_{4}$ & 0.75 \\
    $\alpha \lambda$ & 12.4 \\
    $\Gamma$ & -1.5 \\
    $\mu$ & -0.075 \\ \hline \hline
\end{tabular}
    \caption{Summary of the parameter values used in the Green's function calculation of the dynamical structure factor. }
\label{Parameter:Table}
\end{table}
In the case of a purely tetragonal or trigonal distortion, the sign of the distortion parameter can be inferred from the crystal structure, with $\Gamma<0$ corresponding to a compression of the octahedron and an orbital doublet ground state \cite{Tchernyshyov04:93,Opic57:238}. The term originating from the low symmetry nature of the local environment, appearing in $\mathcal{H}^{\prime}_{dis}$, has no such intuitive interpretation. However, this term fully breaks the degeneracy of the $l=1$ ground state and hence results in an orbital singlet ground state, regardless of the sign of this distortion. We therefore take this distortion to be negative along with the tetragonal distortion. 

\subsection{Neutron scattering response}
The neutron scattering response is plotted in Fig. \ref{Realistic_fig} $(a,b)$ for a tetragonal distortion (Eqn. \ref{tetragonaldistortion}) and Fig. \ref{Realistic_fig} $(d,e)$ for an asymmetric distortion (Eqn. \ref{distortion22}). For both distortion types the spectra are qualitatively similar to the measured neutron response~\cite{Songvilay18:121} with a gapped upper dispersive mode which reaches the zone boundary at around $E \approx$ 10 meV. A further low energy mode is seen at around $E \approx$ 2.5 meV. This mode has a smaller gap and bandwidth, with a spin wave velocity that approaches zero away from the zone center. Both modes are observed to split for this set of parameters, in agreement with Ref. \onlinecite{Songvilay18:121}. The splitting of both modes shows some difference between the two distortions, reflecting the quantitative difference between the Fe$^{2+}$ single-ion energy levels for each distortion. 

The presence of these modes in the linear spin wave calculation for RbFe$^{2+}$Fe$^{3+}$F$_{6}$~\cite{Songvilay18:121} is reflective of the predominant transverse component, which is also captured by the Green's function formalism presented here.  

\subsection{Amplitude fluctuations}

A particular aspect of this analysis is the prediction of amplitude fluctuations in the neutron scattering response to first order in the Dyson expansion where ${{d\langle \hat{S}_{z} \rangle} \over {dt}}\neq 0$.  Such excitations are not present in conventional spin wave theory based on the Landau equation. In this section we analyze the key ingredients that allow such fluctuations to exist to first order in the neutron scattering response. It is important to note that these excitations appear in the $zz$ component of the rotating frame, where the spins are coaligned and fluctuations in the magnitude of the order parameter appear along the common $\hat{z}$-axis. Upon rotating back to the laboratory frame, these fluctuations are no longer confined to the $zz$ component of the structure factor. We shall therefore examine the structure factor in the rotating frame so that the longitudinal, $\tilde{S}^{zz}(\mathbf{q},\omega)$, and transverse components can be distinctly identified. The longitudinal component for both distortions is plotted in Fig. \ref{Realistic_fig} $(c,f)$. With $\Gamma^{\prime} = -1.5$ meV, a weak longitudinal component can be observed (Fig. \ref{Realistic_fig} $(f)$), manifested in a flat mode with $E \approx 17$ meV. The nature of these amplitude modes will now be further investigated.

\subsubsection{Inter-multiplet spin-orbit excitons}
\begin{figure}[h!t]
    \begin{center}
    \includegraphics[trim = {0mm 0mm 0mm 0mm},width=\linewidth]{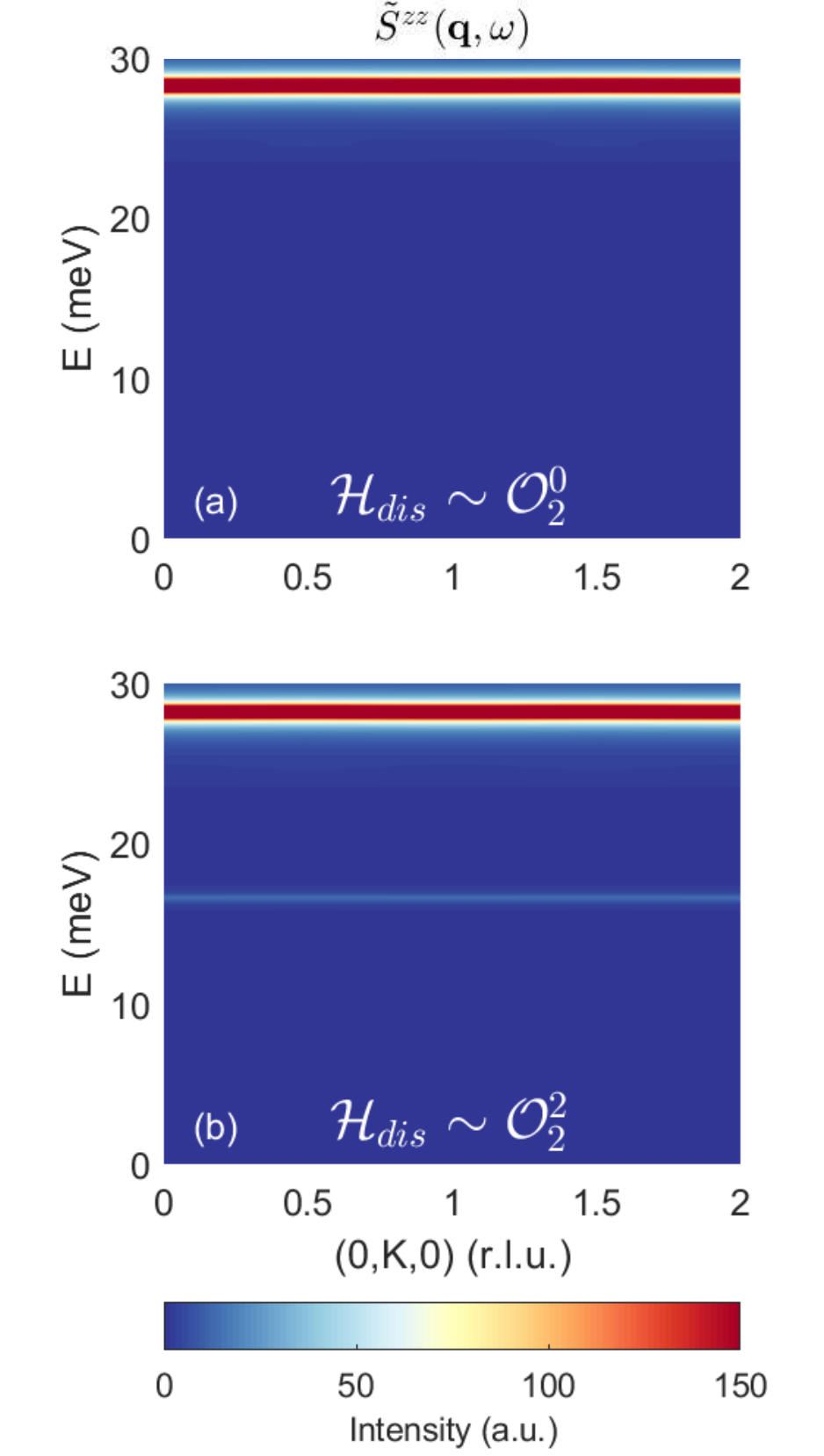}
    \end{center}
    \caption{Spin-orbit exciton at $E\approx$ 28 meV for both types of distortion allowed by symmetry in RbFe$^{2+}$Fe$^{3+}$F$_{6}$. $(a)$ In the case where $\mathcal{H}_{dis}=\mathcal{O}_{2}^{0}$, the spin-orbit exciton is the only amplitude fluctuation that carries non-neglibile intensity. $(b)$ The distortion $\sim \mathcal{O}_{2}^{2}$ exhibits a further flat mode at around 17 meV.  } 
    \label{SOexciton:Fig}
\end{figure}

Regardless of the nature of the distortion, longitudinal transitions between the $j_{eff}=1$ and $j_{eff}=2$ multiplet are permitted (Fig. \ref{LongitudinalTransitionsFig}). These modes generally occur at a higher energy scale than the dispersive magnon excitations, since the energy scale of these excitations are $\sim \lambda$ as per the Land{\'e} interval rule. These modes are particularly susceptible to decay since there is often a large kinematically allowed decay region. The longitudinal component in the rotating frame is plotted in Fig. \ref{SOexciton:Fig} for both of the distortion terms, with $\Gamma=-1.5$ meV and $\Gamma^{\prime}=-1.5$ meV respectively. For each of these distortions, a high energy spin-orbit exciton is seen at $E \approx$ 28 meV. 

\subsubsection{Intra-multiplet distortion modes}

We now turn our attention to intra-multiplet modes. In Fig. \ref{SOexciton:Fig}, $(b)$ a second flat mode can be seen at $E \approx$ 17 meV, originating from the intra-multiplet transition which gains intensity under an asymmetrically distorted crystal field. This mode is weak, in agreement with Fig. \ref{LongitudinalTransitionsFig}, which suggests that the intensity of this amplitude mode is suppressed by the molecular field. It should also be noted that this mode is likely susceptible to decay owing to the fact that it lies at an energy that is less than two times the magnitude of the expected magnon bandwidth~\cite{Songvilay18:121}.

 \begin{figure*}[h!t]
    \begin{center}
    \includegraphics[trim = {0mm 0mm 0mm 0mm},width=\linewidth]{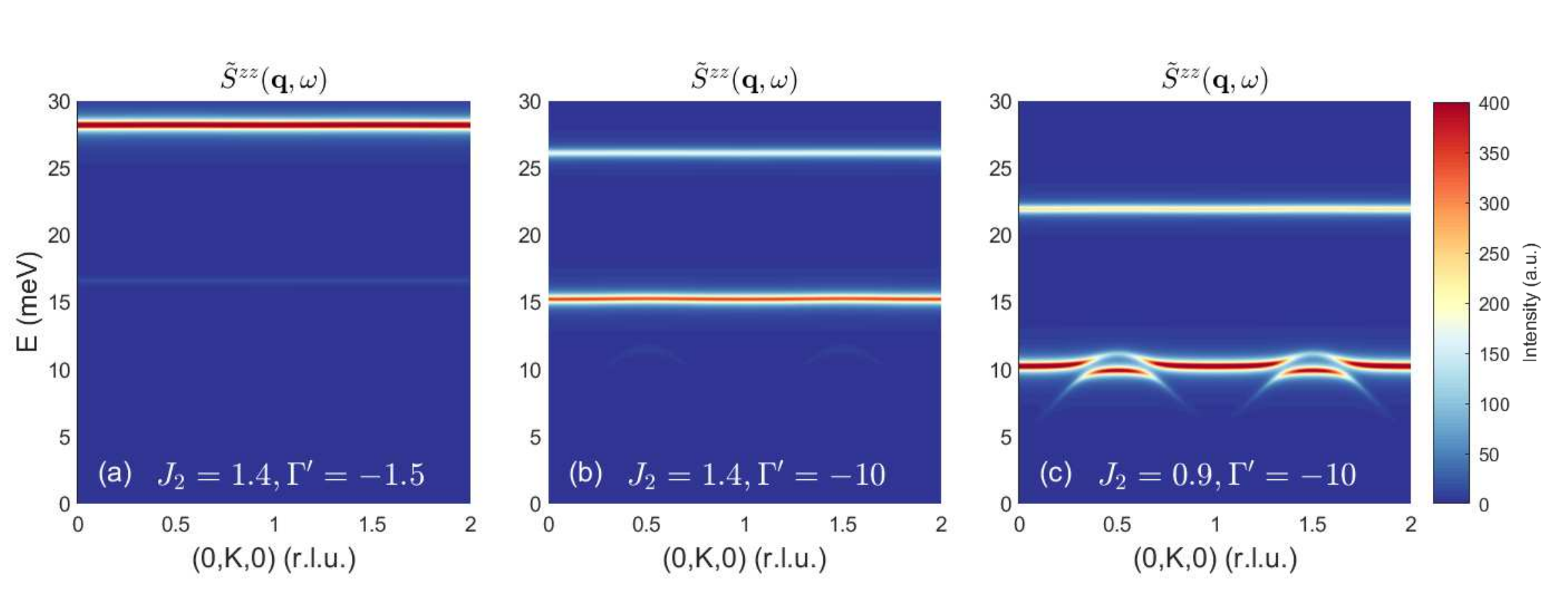}
    \end{center}
    \caption{$(a)$ Longitudinal component of the structure factor in the rotating frame for $J_{2}=1.4$ meV and $\Gamma^{\prime}=-1.5$ meV. Both the high energy spin-orbit exciton and a flat weak intra-multiplet mode are visible. $(b)$ Amplitude fluctuations with $J_{2}=1.4$ meV and $\Gamma^{\prime}=-10$ meV. Upon increasing the magnitude of the distortion, the flat intra-multiplet mode gains intensity. A very weak dispersive lower mode appears around $E\approx$ 12 meV. $(c)$ Upon decreasing the molecular mean field $h_{MF}$ by decreasing $J_{2}$, the lower two modes hybridize and the intra-multiplet mode increases in intensity.  } 
    \label{DistortionSFact:Fig}
\end{figure*}
We now further explore the nature of the asymmetric distortion. Fig. \ref{DistortionSFact:Fig} $(a)$ shows the longitudinal component of the structure factor for the asymmetric distortion $\mathcal{H}_{dis}^{\prime}$, with $\Gamma^{\prime}=-1.5$ meV. A weak flat mode at $E \approx$ 17 meV is visible. Upon increasing the magnitude of the distortion parameter to $\Gamma^{\prime}=-10$ meV, the flat mode gains spectral weight and a very weak dispersive amplitude mode at lower energy appears (Fig. \ref{DistortionSFact:Fig} $(b)$). Finally, after reducing the magnitude of $J_{2}$ and hence $|h_{MF}|$, the intensity of the intra-multiplet modes can be seen to increase in Fig. \ref{DistortionSFact:Fig} $(c)$ (in agreement with Fig. \ref{LongitudinalTransitionsFig}) and the flat intra-multiplet mode hybridizes with the lower dispersive mode.  

\section{Discussion and concluding remarks}

We have presented an excitonic description of the spin excitations in insulating RbFe$^{2+}$Fe$^{3+}$F$_{6}$ applying a multi-level formalism with Green's functions.  This approach differs from semiclassical descriptions which focus on transverse perturbations of a spin of fixed magnitude.  While such approaches incorporate local anisotropy through anisotropic and antisymmetric terms, the Greens function approach applied here explicitly incorporates single-ion physics and spin-orbit coupling.  Bringing in spin-orbit coupling ($\propto\mathbf{l}\cdot \mathbf{S}$) is particularly important as the observable operator $\hat{S}_{z}$ no longer commutes with the Hamiltonian $[\mathcal{H},\hat{S}_{z}]\neq0$ and therefore the expectation value $\langle \hat{S}_{z} \rangle$ is no longer explicitly a conserved quantity (implying ${{d\langle \hat{S}_{z} \rangle} \over {dt}}\neq 0$).  This allows unusual types of excitations such as amplitude fluctuations to become allowed and observable with the dipolar selection rules of neutron scattering and also optical techniques such as Raman.  As discussed above, such excitations are no longer forbidden in RbFe$^{2+}$Fe$^{3+}$F$_{6}$ owing to the presence of an orbitally degenerate ground state of Fe$^{2+}$ (as schematically illustrated in Fig. \ref{SingleionFig}).

One of the issues with experimentally observing amplitude modes resulting from excitonic magnetic excitations is that they typically occur at higher energies than the lower energy transverse excitations.  Typically, these modes then decay and appear experimentally as an energy and momentum broadened continuum of scattering, not a temporally sharp underdamped excitation like a harmonic spin wave or a sharp dispersionless crystal field excitation.  Such a situation has been analyzed theoretically and experimentally in the fourth row transition metal ion compound Ca$_{2}$RuO$_{4}$~\cite{Sarte20:102,Jain13:17}. In this particular situation the amplitude mode was kinematically allowed to decay into lower transverse modes resulting in a continuum of scattering observable with the combination of polarized neutrons and the mapping capabilities afforded by modern neutron spectrometers.  We note that given the formalism presented here, and applied in Ref. \onlinecite{Sarte20:102}, only corresponds to first order mean field theory, it does not capture such decay process which require higher order terms in the Dyson expansion.  This is beyond the scope and the goal of the analysis presented here.

In this context, it is interesting, to apply this to the case of RbFe$^{2+}$Fe$^{3+}$F$_{6}$.  As experimentally reported in Ref. \onlinecite{Songvilay18:121}, the magnetic excitations consist of two components -- a temporally well-defined underdamped component and also a component that is broadened in both energy and momentum.  Such a component may originate from quantum fluctuations owing to noncommuting observables reported in low spin chains, however it is not expected to be strong in large spin components such as $S=2$ of Fe$^{2+}$ or $S={5\over 2}$ of Fe$^{3+}$.  This leads us to suggest in this paper that it may originate from amplitude fluctuations allowed by the low local symmetry of the Fe$^{2+}$ ion and the presence of spin-orbit coupling.     

This work illustrates that there are two components required required for the presence of observable amplitude fluctuations at accessible low-energies in intermediate field third-row transition metal ions.  The first is spin-orbit coupling like found here in Fe$^{2+}$ or present in V$^{3+}$ or Co$^{2+}$ which allows fluctuations in the order parameter amplitude $\langle \hat{S}_{z} \rangle$ to occur.  The second key component is the presence of low symmetry, permitting single-ion terms such as $\sim \mathcal{O}_{2}^{2}$ which are not present in tetragonal, trigonal or hexagonal symmetry~\cite{Bauer:book}. A distortion of this form can enhance the intensity of amplitude fluctuations, both in the form of spin-orbit excitons ($\sim \lambda$) and lower energy intra-multiplet modes which can disperse. While tetragonal distortions can give rise to amplitude fluctuations, these are typically present at higher energies close to the single-ion energy scale of the spin-orbit transitions ($\sim \lambda$), which is $\sim$ 30 meV in third row transition metal ions. Such fluctuations are less relevant as it is much more difficult to tune to such energy scales or stabilize them.  Therefore, it is suggested that amplitude modes in third row transition metal ions are best sought in compounds with low local symmetry and based on magnetic ions with an orbital degeneracy like Fe$^{2+}$, V$^{3+}$, or Co$^{2+}$.  
\begin{acknowledgements}
The authors thank W. J. L. Buyers, P. M. Sarte, N. Giles-Donovan, C. Batista and B. Roessli for useful discussions. Experiments at the ISIS Pulsed Neutron and Muon Source were supported by beamtime allocation RB1620320 from the Science and Technology Facilities Council. H. L. was co-funded by the ISIS facility development studentship programme. This work was supported by the EPSRC and the STFC.
\end{acknowledgements}
\bibliography{RbFeF_references}
\appendix
\section{Definition of exchange interaction}
\label{AppendixA}
Central to the calculation is the Fourier transform of the exchange interaction, $\underline{\underline{\mathcal{J}}}(\mathbf{q})=\sum_{ij}\underline{\underline{\mathcal{J}}}_{ij}^{\gamma\gamma'}e^{i\mathbf{q}\cdot (\mathbf{r}_{i\gamma}-\mathbf{r}_{j\gamma'})}$. The matrix elements can be calculated using Table \ref{Bonds}, the nonzero elements are listed below
\begin{widetext}
\begin{align*}
    \left[\underline{\underline{\mathcal{J}}}(\mathbf{q})\right]_{12}=&J_{1}\left(e^{i\mathbf{q}\cdot (\mathbf{r}_{2}-\mathbf{r}_{1})}+e^{i\mathbf{q}\cdot (\mathbf{r}_{2}-\mathbf{r}_{1}-[0,1,0])}\right)\\
     \left[\underline{\underline{\mathcal{J}}}(\mathbf{q})\right]_{15}=&J_{3}e^{i\mathbf{q}\cdot(\mathbf{r}_{5}-\mathbf{r}_{1}-[0,1,0])}\\
    \left[\underline{\underline{\mathcal{J}}}(\mathbf{q})\right]_{16}=&J_{4}e^{i\mathbf{q}\cdot(\mathbf{r}_{6}-\mathbf{r}_{1}-[1,1,0])}\\
    \left[\underline{\underline{\mathcal{J}}}(\mathbf{q})\right]_{17}=&J_{4}e^{i\mathbf{q}\cdot(\mathbf{r}_{7}-\mathbf{r}_{1}-[0,0,1])}\\
         \left[\underline{\underline{\mathcal{J}}}(\mathbf{q})\right]_{18}=&J_{3}e^{i\mathbf{q}\cdot(\mathbf{r}_{8}-\mathbf{r}_{1}-[1,0,1])}\\
    \left[\underline{\underline{\mathcal{J}}}(\mathbf{q})\right]_{25}=&J_{3}e^{i\mathbf{q}\cdot(\mathbf{r}_{5}-\mathbf{r}_{2})}\\
   \left[\underline{\underline{\mathcal{J}}}(\mathbf{q})\right]_{26}=&J_{4}e^{i\mathbf{q}\cdot(\mathbf{r}_{6}-\mathbf{r}_{2}-[1,0,0])}\\
   \left[\underline{\underline{\mathcal{J}}}(\mathbf{q})\right]_{27}=&J_{4}e^{i\mathbf{q}\cdot(\mathbf{r}_{7}-\mathbf{r}_{2}-[0,0,1])}\\
    \left[\underline{\underline{\mathcal{J}}}(\mathbf{q})\right]_{28}=&J_{3}e^{i\mathbf{q}\cdot(\mathbf{r}_{8}-\mathbf{r}_{2}-[1,0,1])}\\
    \left[\underline{\underline{\mathcal{J}}}(\mathbf{q})\right]_{34}=&J_{1}\left(e^{i\mathbf{q}\cdot(r_{4}-r_{3})}+e^{i\mathbf{q}\cdot(r_{4}-r_{3}-[0,1,0])}\right)\\
    \left[\underline{\underline{\mathcal{J}}}(\mathbf{q})\right]_{35}=&J_{4}e^{i\mathbf{q}\cdot(\mathbf{r}_{5}-\mathbf{r}_{3}-[0,1,0])}\\
   \left[\underline{\underline{\mathcal{J}}}(\mathbf{q})\right]_{36}=&J_{3}e^{i\mathbf{q}\cdot(\mathbf{r}_{6}-\mathbf{r}_{3}-[0,1,0])}\\
     \left[\underline{\underline{\mathcal{J}}}(\mathbf{q})\right]_{37}=&J_{3}e^{i\mathbf{q}\cdot(\mathbf{r}_{7}-\mathbf{r}_{3})}\\
      \left[\underline{\underline{\mathcal{J}}}(\mathbf{q})\right]_{38}=&J_{4}e^{i\mathbf{q}\cdot(\mathbf{r}_{8}-\mathbf{r}_{3})}\\
    \left[\underline{\underline{\mathcal{J}}}(\mathbf{q})\right]_{45}=&J_{4}e^{i\mathbf{q}\cdot(\mathbf{r}_{5}-\mathbf{r}_{4})}\\
  \left[\underline{\underline{\mathcal{J}}}(\mathbf{q})\right]_{46}=&J_{3}e^{i\mathbf{q}\cdot(\mathbf{r}_{6}-\mathbf{r}_{4})}\\
    \left[\underline{\underline{\mathcal{J}}}(\mathbf{q})\right]_{47}=&J_{3}e^{i\mathbf{q}\cdot(\mathbf{r}_{7}-\mathbf{r}_{4})}\\
  \left[\underline{\underline{\mathcal{J}}}(\mathbf{q})\right]_{48}=&J_{4}e^{i\mathbf{q}\cdot(\mathbf{r}_{8}-\mathbf{r}_{4})}\\
   \left[\underline{\underline{\mathcal{J}}}(\mathbf{q})\right]_{56}=&J_{2}\left(e^{i\mathbf{q}\cdot(\mathbf{r}_{6}-\mathbf{r}_{5})}+e^{i\mathbf{q}\cdot(\mathbf{r}_{6}-\mathbf{r}_{5}-[1,0,0])}\right)\\
    \left[\underline{\underline{\mathcal{J}}}(\mathbf{q})\right]_{78}=&J_{2}\left(e^{i\mathbf{q}\cdot(\mathbf{r}_{8}-\mathbf{r}_{7})}+e^{i\mathbf{q}\cdot(\mathbf{r}_{8}-\mathbf{r}_{7}-[1,0,0])}\right).
\end{align*}
\end{widetext}
\noindent The corresponding elements in the lower left triangle can be found by reversing position vector labels, hence the matrix is Hermitian. 
\end{document}